\documentclass[fleqn,usenatbib]{mnras}

\usepackage{newtxtext,newtxmath}
\usepackage[T1]{fontenc}

\DeclareRobustCommand{\VAN}[3]{#2}
\let\VANthebibliography\thebibliography
\def\thebibliography{\DeclareRobustCommand{\VAN}[3]{##3}\VANthebibliography}

\usepackage{graphicx}	% Including figure files
\usepackage{amsmath}	% Advanced maths commands
\usepackage{gensymb}
\usepackage{hyperref}

\newcommand{\change}[1]{#1}
\newcommand{\changetwo}[1]{#1}
\newcommand{\mj}{MAXI~J1803$-$298}
\newcommand{\perhour}{hr$^{-1}$}

\title[\mj\ time-dependent visibility modelling]{Time-dependent visibility modelling of a relativistic jet in the X-ray binary \mj}

\author[C. M. Wood et al.]{
C. M. Wood,$^{1}$\thanks{E-mail: callan.wood@icrar.org}
J. C. A. Miller-Jones,$^{1}$
A. Bahramian,$^{1}$
S. J. Tingay,$^{1}$
T. D. Russell,$^{2}$
A. J. Tetarenko,\thanks{NASA Einstein Fellow}$^{3,4}$\newauthor
D. Altamirano,$^{5}$
T. Belloni,$^{6}$
F. Carotenuto,$^{7}$
C. Ceccobello,$^{8}$
S. Corbel,$^{9}$
M. Espinasse,$^{9}$
R. P. Fender,$^{7,10}$\newauthor
E. K{\"o}rding,$^{11}$
S. Migliari,$^{12,13}$
D. M. Russell,$^{14}$
C. L. Sarazin,$^{15}$
G. R. Sivakoff,$^{16}$
R. Soria,$^{17,18,19}$
and \newauthor
V. Tudose.$^{20}$
\\
$^{1}$International Centre for Radio Astronomy Research, Curtin University, GPO Box U1987, Perth, WA 6845, Australia\\
$^{2}$INAF, Istituto di Astrofisica Spaziale e Fisica Cosmica, Via U. La Malfa 153, I-90146 Palermo, Italy\\
$^{3}$Department of Physics and Astronomy, Texas Tech University, Lubbock, TX 79409-1051, USA\\
$^{4}$East Asian Observatory, 660 N. A`oh\={o}k\={u} Place, University Park, Hilo, Hawaii 96720, USA\\
$^{5}$School of Physics and Astronomy, University of Southampton, Southampton, Hampshire SO17 1BJ, UK\\
$^{6}$INAF, Osservatorio Astronomico di Brera Via E. Bianchi 46, I-23807 Merate, Italy\\
$^{7}$Astrophysics, Department of Physics, University of Oxford, Keble Road, Oxford OX1 3RH, UK\\
$^{8}$Department of Space, Earth and Environment, Chalmers University of Technology, Onsala Space Observatory, SE-439 92 Onsala, Sweden\\
$^{9}$Universit{\'e} Paris Cit{\'e} and Universit{\'e} Paris Saclay, CEA, CNRS, AIM, F-91190 Gif-sur-Yvette, France\\
$^{10}$Department of Astronomy, University of Cape Town, Private Bag X3, Rondebosch 7701, South Africa\\
$^{11}$Department of Astrophysics/IMAPP, Radboud University, PO Box 9010, NL-6500 GL Nijmegen, the Netherlands\\
$^{12}$Aurora Technology for the European Space Agency, ESAC/ESA, Camino Bajo del Castillo s/n, Urb. Villafranca del Castillo, \\\quad 28691 Villanueva de la Cañada, Madrid, Spain\\
$^{13}$Institut de Ciències del Cosmos (ICC), Universitat de Barcelona (IEEC-UB), Martí i Franquès 1, 08028 Barcelona, Spain\\
$^{14}$Center for Astro, Particle and Planetary Physics, New York University Abu Dhabi, PO Box 129188, Abu Dhabi, UAE\\
$^{15}$Department of Astronomy, University of Virginia, 530 McCormick Road, Charlottesville, VA 22904-4325, USA\\
$^{16}$Faculty of Science, University of Alberta, CCIS 4-181, Edmonton, AB T6G 2E1, Canada\\
$^{17}$INAF, Osservatorio Astrofisico di Torino, Strada Osservatorio 20, 10025, Pino Torinese, Italy\\
$^{18}$College of Astronomy and Space Sciences, University of the Chinese Academy of Sciences, Beijing 100049, China\\
$^{19}$Sydney Institute for Astronomy, School of Physics A28, The University of Sydney, Sydney, NSW 2006\\
$^{20}$Institute for Space Sciences, Atomistilor 409, PO Box MG-23, 077125 Bucharest-Magurele, Romania}
\date{Accepted XXX. Received YYY; in original form ZZZ}

\pubyear{2023}

\begin{document}
\label{firstpage}
\pagerange{\pageref{firstpage}--\pageref{lastpage}}
\maketitle

\begin{abstract}
    Tracking the motions of transient jets launched by low-mass X-ray binaries (LMXBs) is critical for determining the moment of jet ejection, and identifying any corresponding signatures in the accretion flow. However, these jets are often highly variable and can travel across the resolution element of an image within a single observation, violating a fundamental assumption of aperture synthesis. We present a novel approach in which we directly fit a single time-dependent model to the full set of interferometer visibilities, where we explicitly parameterise the motion and flux density variability of the emission components, to minimise the number of free parameters in the fit, while leveraging information from the full observation. This technique allows us to detect and characterize faint, fast-moving sources, for which the standard time binning technique is inadequate. We validate our technique with synthetic observations, before applying it to three Very Long Baseline Array (VLBA) observations of the black hole candidate LMXB \mj\ during its 2021 outburst. We measured the proper motion of a discrete jet component to be $1.37\pm0.14$~mas\,\perhour, and thus we infer an ejection date of MJD~$59348.08_{-0.06}^{+0.05}$, which occurs just after the peak of a radio flare observed by the Australia Telescope Compact Array (ATCA) and the Atacama Large Millimeter/Sub-Millimeter Array (ALMA), while \mj\ was in the intermediate state. Further development of these new VLBI analysis techniques will lead to more precise measurements of jet ejection dates, which, combined with dense, simultaneous multi-wavelength monitoring, will allow for clearer identification of jet ejection signatures in the accretion flow.
\end{abstract}


%%%%%%%%%%%%%%%%%%%%%%%%%%%%%%%%%%%%%%%%%%%%%%%%%%

%%%%%%%%%%%%%%%%% BODY OF PAPER %%%%%%%%%%%%%%%%%%

\section{Introduction}
    
    Relativistic jets launched by black holes are some of the most powerful phenomena in the Universe. By studying jets from low-mass X-ray binaries (LMXBs) within our own Galaxy, we can try to understand how these jets are launched. Together with contemporaneous X-ray observations we can then seek to determine the nature of the relationship between the in-flowing accretion material and the formation of these jets. 
    Observations of black hole LMXBs, which are systems consisting of a stellar-mass black hole accreting matter from a low-mass companion star, have identified a number of accretion states. These states correspond to different configurations of the inflowing and outflowing material, and different X-ray and radio spectral and variability signatures. LMXBs spend most of their time in quiescence, interspersed by occasional bright outbursts, which typically consist of a transition from the rising hard state to the soft state, via intermediate states, followed by a decline and reverse transition back into the hard state (see e.g. \citealp{Homan2005TheStates}, and for a review, see \citealp{BelloniMotta2011BlackHoleTransients} and references therein). A typical feature of the hard state is the presence of strong, steady, compact synchrotron-emitting jets. At some point during the state transition, these pre-existing, steady jets are quenched and often, discrete, transient jet ejecta are launched \citep[e.g.][]{MirabelandRodriguez1994AGalaxy, Tingay1995RelativisticSource, Hjellming1995Episodic40, MillerJones2012DiscjetH1743322}, which can travel with apparent superluminal motion out to large distances \citep[e.g.][]{Bright2020AnJ1820+070}. These transient jets are not seen in the subsequent reverse transition back to the hard state \citep{Fender2004TowardsJets, Corbel2004StatesXTE1650}. 
    
    The nature of the causal connection between changes in the inner accretion flow and the ejection of transient jets during the state transition has been the focus of many recent studies. Particular X-ray spectral and timing properties that characterise the evolving accretion flow have been suggested as signatures of jet ejection \citep[e.g.][]{Fender2009JetsXrays, MillerJones2012DiscjetH1743322, Russell2019DiskJetJ1535571, Homan2020AJ1820+070, Mendez2022GRS1915coupling}. However, the nature of the association between such signatures and the moment of jet launching is often unclear \citep[e.g.][]{MillerJones2012DiscjetH1743322}. This is due in part to the scarcity of high angular resolution observations of these transient jet ejecta, in combination with the difficulty of measuring the proper motions of jets that exhibit either significant intra-observational variability, or are only detected in a single observation. To clearly identify the specific signatures of changes in the inner accretion flow associated with the launching of transient relativistic jets, we require precise measurements of their ejection dates, accompanied by simultaneous X-ray observations \citep[e.g.][]{Wood2021Varying}. The angular resolution required to make precise proper motion measurements is only possible with very long baseline interferometry (VLBI).
    
    \subsection{Very Long Baseline Interferometry}
    
        In an interferometer, each pair of telescopes measures a complex visibility, which, according to the van Cittert-Zernike theorem \citep{vanCittert1934DieEbene, Zernike1938TheProblems} is an element of the Fourier transform of the sky brightness distribution. Typical VLBI observations can be up to several hours in length, so that as the Earth rotates, the orientations of the separation vectors (the baselines) of the pairs of telescopes projected onto the plane of the sky change. This samples more unique visibilities, increasing the overall sensitivity and quality of the image reconstruction.
        
        Since the complex visibility plane (often called the $uv$-plane) can never be completely sampled, the inverse Fourier transform of the visibilities is a convolution of the true sky brightness distribution and the inverse Fourier transform of the sampling function. Many imaging algorithms have been developed with the aim of reconstructing the true sky brightness distribution from the incomplete information. In radio astronomy, the standard technique is the CLEAN algorithm \citep{Hogbom1974ApertureBaselines, Schwarz1978Mathematical-statisticalCLEAN, Clark1980AnCLEAN}, which is a deconvolution technique that represents the sky as a sum of point sources, and attempts to iteratively subtract out the artefacts and side lobes of the sampling function from the inverse Fourier transform of the visibilities.
        
        There is another class of techniques that first attempt to reconstruct a version of the sky brightness distribution, before comparing that reconstruction to the underlying data. Examples of this class of methods are the so called maximum entropy methods (MEM) or regularised maximum likelihood (RML) methods, that try to solve for the best possible image by fitting the image pixels to the data while providing constraints via the use of regularization terms that favour certain features in the image, e.g. entropy, sparsity, or smoothness \citep{Frieden1972MEM, Cornwell1985Mem, Narayan1986MaximumAstronomy}. These methods are not as popular as the CLEAN algorithm, although in recent years they have gained attention, particular by groups such as the Event Horizon Telescope (EHT) collaboration \citep[e.g.][]{Chael2016HIGH-RESOLUTIONTELESCOPE, Chael2018InterferometricAmplitudes, Akiyama2017imaging, EHTCollaboration20194, Broderick2020ThemisHyrbid,Broderick2022variabilitymitigate}. 
        
        Another example of this approach is model fitting. In this technique, simple model source components with analytic representations in the Fourier domain (e.g. point sources or Gaussians) are fit directly to the visibilities \citep[e.g.][]{Shepherd1994DIFMAP, Marti-Vidal2014UVMMULTIFIT}, greatly reducing the number of free parameters in the imaging problem. Historically, model fitting was first used before imaging, with early two element interferometers \citep[e.g.][]{Fomalont1968eastweststructure}. Model fitting has been used to study transient jets launched by LMXBs, since they are often seen in images as compact point sources or Gaussians \citep[e.g.][]{MillerJones2019V404}. 
        
        One of the fundamental assumptions of VLBI is that over the length of an observation the target source is non-variable. Jets launched by LMXBs can travel across the resolution element of an image in a matter of minutes \citep[e.g.][]{Wood2021Varying} and can vary by a significant fraction of their flux density on the same time-scale \citep[e.g.][]{MillerJones2019V404}, violating this assumption. The simplest solution is time binning \citep[e.g.][]{Fomalont2001ScorpiusLobes, MillerJones2019V404}, where the full observation is split into short time bins, within which the source is relatively static, each to be imaged individually. This technique requires the source to be bright enough so that it can be significantly detected in each time bin, since within a single time bin the sensitivity and $uv$-coverage are greatly reduced, making this technique difficult for standard LMXB ejecta, which are typically only tens of mJy in brightness. 
        
        Recently, more sophisticated techniques have been developed that seek to improve upon the time binning procedure. In \citet{Wood2021Varying} we described a dynamic phase centre tracking technique by which we applied an incremental phase shift to each time bin of an observation before stacking the time bins back together to effectively 'track' a jet component with a given proper motion. Other recent developments have been focused on capturing the variability  of the super-massive black holes M87$^*$ and Sagittarius A$^*$ in EHT observations \citep[e.g.][]{bouman2017reconstructing, Johnson2017DynamicalInterferometry, Arras2022VariableM87}. These approaches aim to extend MEM and RML methods to simultaneously reconstruct images from all of the time bins in an observation, while explicitly enforcing continuity across the full set of images, to leverage information from the entire observation to enhance the quality and sensitivity of each individual image. The further development of techniques that can capture intra-observational variability is key to making more precise measurements of the proper motions and ejection dates of transient jets, and thus determining the causal connection between changes in the inner accretion flow and the launching of relativistic jets in LMXBs.
        
    \subsection{\mj\ }
        \mj\ (hereafter J1803) was first discovered as a new X-ray transient in the early stages of an outburst on \change{2021 May 1} (MJD 59335) by the Monitor of All-sky X-ray Image \citep[MAXI;][]{Serino2021J1803ATel} nova alert system. It was quickly localised by both NICER \citep{Gendreau2021NICERJ1803, Gendreau2021ATelcorreection} and \textit{Swift}, the latter of which also detected an optical counterpart \citep{Gropp2021swfitJ1803}. Spectroscopy with the Southern African Large Telescope suggested that J1803 was an LMXB \citep{Buckley2021ATelSALTJ1803}. NICER, \textit{AstroSat}, and \textit{NuSTAR} X-ray spectral analysis further suggested that J1803 was an accreting stellar-mass black hole, as opposed to an accreting neutron star \citep[][]{Bult2021ATelNICERJ1803, Jana2021ATelAstroSATJ1803, Xu2021ATelNuSTARJ1803}, viewed relatively edge-on, with an inclination above $70\degree$. On \change{2021 May 4} (MJD 59338.9), \citet{Espinasse2021ATelJ1803} first detected J1803 at radio wavelengths with MeerKAT. On \change{2021 May 11} (MJD 59345), 10 days after its initial discovery, \textit{AstroSat} detected a state transition of J1803 \citep{Jana2021ATelAstroSATJ1803}, with their observations suggesting that the source had entered the hard-intermediate state. MAXI/GSC was unable to observe J1803 for eight days from \change{2021 May 4} (MJD 59338), but on \change{2021 May 12} (MJD 59346) they also reported that the source was in the intermediate state, with the transition to the soft state occurring on \change{2021 May 28} \citep[MJD 59362;][]{ShidatsuJ18032022}. J1803 remained in the soft state for $\sim5$ months, with the reverse soft-to-hard state transition occurring between \change{2021 October 13 and 2021 October 19} \citep[MJD 59500-59506;][]{Steiner2021J1803ATel}.
        
        We present the results of a radio monitoring campaign of J1803 during the state transition, with the Very Long Baseline Array (VLBA), the Australia Telescope Compact Array (ATCA), and the Atacama Large Millimeter/Sub-Millimeter Array (ALMA). In order to account for the intra-observational variability of our VLBA observations, we have developed a new model-fitting approach in which we jointly fit a single time-evolving model to all of the visibilities in a single observation, rather than on a time bin by time bin basis. This allowed us to leverage all of the information from a full observation in a single fit to constrain the motion and flux density variability of the detected components. We first demonstrate our validation of this technique with synthetic data sets designed to replicate the typical variability we would expect in our observations, before presenting the results of our application of this technique to our VLBA observations of J1803. 
        
        The paper is organised as follows. We describe our observations, calibration, imaging, and model fitting procedure in Section~\ref{sec:methods}. We present the results of this analysis in Section~\ref{sec:results}. We discuss our results in Section~\ref{sec:discussion} and present our conclusions in Section~\ref{sec:conclusion}.

\section{Methods} \label{sec:methods}
    \subsection{Observations and Calibration}
        \subsubsection{VLBA}
            Following the initial X-ray detection of the outburst \citep{Serino2021J1803ATel}, we observed J1803 with the VLBA as part of the Jet Acceleration and Collimation Probe Of Transient X-Ray Binaries \citep[JACPOT XRB;][]{JACPOTMillerJones} program. We conducted 15 observations between \change{2021 May 13} and \change{2021 June 7}, around and following the peak of the outburst and the state transition. To better characterise the intra-day motions of the evolving jets, nine of these observations were split into two short (1-hour) blocks separated by $\sim1$ hour. We also conducted an astrometric observation on \change{2021 November 12}, following the transition back into the hard state \citep{Steiner2021J1803ATel}. The details of the observations are listed in Table~\ref{tab:VLBA Observation Log}. 
            
            In epoch A we observed in X-band (8.4\,GHz) with a recording rate of 2048 Mbps, with a total bandwidth of 256 MHz split into eight 32-MHz intermediate-frequency (IF) pairs. In the subsequent epochs we observed in the wide-band mode, with a recording rate of 4096 Mbps, yielding a total bandwidth of 512 MHz split into four 128-MHz IF pairs. In epochs B1 to I2, we observed at 8.3\,GHz. In epochs J1 to O, we observed using the dual S/X-band dichroic feed, with the first IF pair containing the S-band (2.2\,GHz) data and the other three IF pairs containing the X-band (8.2\,GHz) data. For these observations we split the data into the two separate bands to calibrate and image separately. In the final observation, epoch P, we observed in the most sensitive C-band (4.9\,GHz), aiming to detect J1803 as it faded into quiescence. Alongside the target source, we observed J1743-0350 and 1921-293 as fringe finders, J1803-2748 \citep{Shu2017} as a phase reference calibrator, and J1752-2956 \citep{Petrov2006} as a check source. \changetwo{In all observations, we observed in dual circular polarisation mode, combining RR and LL correlations to create Stokes I. Due to the short duration of the observing blocks, we did not set the observations up for polarisation calibration, and were thus unable to reliably measure Stokes Q, U, or V.}
            
            In epochs A and P, we observed geodetic blocks \citep{Reid2009Geodetic} for $\sim$20 minutes at the beginning and end of the observations to correct for unmodelled tropospheric delays and clock errors. The data were correlated using the DiFX software correlator \citep{Deller2007DiFX1,Deller2011DiFX-2:Correlator}, and calibrated following the standard procedures within the Astronomical Image Processing System \citep[\textsc{aips}, version 31DEC22;][]{Wells1985NRAOsAIPS,Greisen2003AIPSVLBA}. \change{Following the standard external gain calibrations, we performed several rounds of hybrid mapping with the phase reference calibrator to make the best possible model, to derive the time-varying phase, delay, and rate solutions, which we interpolated to the target source. We also applied the amplitude gain solutions from the hybrid mapping of the phase reference source to the target source, to get most accurate time-varying amplitude gain calibration.} 
        
            \begin{table*}
                \centering
                \caption{VLBA observation log for the 2021 JACPOT campaign on \mj, under project code BM509. \changetwo{MJD denotes the mid-time of each observation.}}
                \label{tab:VLBA Observation Log}
                \begin{tabular}{|c|c|c|c|c|c|c|}
                \hline
                Epoch & Date      & MJD               & Time        & Frequency & Bandwidth & Spectral State$^\dag$ \\ 
                      &    &                   & (UTC)       & (GHz)     & (MHz)     & \\ 
                \hline \hline
                A     &  \change{2021 May 13} & \change{59347.41} & 07:45-11:44 & 8.4       & 256       & Intermediate \\
                B1    &  \change{2021 May 14} & \change{59348.36} & 08:13-09:13 & 8.3       & 512       & Intermediate \\
                B2    &  \change{2021 May 14} & \change{59348.47} & 10:42-11:42 & 8.3       & 512       & Intermediate \\
                C     &  \change{2021 May 15} & \change{59349.41} & 08:54-10:53 & 8.3       & 512       & Intermediate \\
                D     &  \change{2021 May 16} & \change{59350.44} & 09:35-11:34 & 8.3       & 512       & Intermediate \\
                E1    &  \change{2021 May 18} & \change{59352.34} & 07:42-08:42 & 8.3       & 512       & Intermediate \\
                E2    &  \change{2021 May 18} & \change{59352.45} & 10:12-11:12 & 8.3       & 512       & Intermediate \\
                F1    &  \change{2021 May 19} & \change{59353.33} & 07:23-08:23 & 8.3       & 512       & Intermediate \\
                F2    &  \change{2021 May 19} & \change{59353.45} & 10:23-11:23 & 8.3       & 512       & Intermediate \\
                G     &  \change{2021 May 21} & \change{59355.44} & 09:30-11:30 & 8.3       & 512       & Intermediate \\
                H1    &  \change{2021 May 22} & \change{59356.33} & 07:26-08:26 & 8.3       & 512       & Intermediate \\
                H2    &  \change{2021 May 22} & \change{59356.41} & 09:26-10:26 & 8.3       & 512       & Intermediate \\
                I1    &  \change{2021 May 23} & \change{59357.35} & 07:52-08:52 & 8.3       & 512       & Intermediate \\
                I2    &  \change{2021 May 23} & \change{59357.43} & 09:52-10:52 & 8.3       & 512       & Intermediate \\
                J1$^{\ddag}$    &  \change{2021 May 28} & \change{59362.30} & 06:48-07:48 & 2.2 / 8.2   & 128 / 384   & Soft \\
                J2$^{\ddag}$    &  \change{2021 May 28} & \change{59362.43} & 09:47-10:47 & 2.2 / 8.2   & 128 / 384   & Soft \\
                K1$^{\ddag}$    &  \change{2021 May 30} & \change{59364.34} & 07:40-08:40 & 2.2 / 8.2   & 128 / 384   & Soft \\
                K2$^{\ddag}$    &  \change{2021 May 30} & \change{59364.42} & 09:39-10:39 & 2.2 / 8.2   & 128 / 384   & Soft \\
                L1$^{\ddag}$    &  \change{2021 May 31} & \change{59365.31} & 06:51-07:51 & 2.2 / 8.2   & 128 / 384   & Soft \\
                L2$^{\ddag}$    &  \change{2021 May 31} & \change{59365.39} & 08:51-09:50 & 2.2 / 8.2   & 128 / 384   & Soft \\
                M1$^{\ddag}$    &  \change{2021 June 01} & \change{59366.29} & 06:32-07:32 & 2.2 / 8.2   & 128 / 384   & Soft \\
                M2$^{\ddag}$    &  \change{2021 June 01} & \change{59366.38} & 08:32-09:32 & 2.2 / 8.2   & 128 / 384   & Soft \\
                N$^{\ddag}$     &  \change{2021 June 03} & \change{59368.35} & 06:24-10:24 & 2.2 / 8.2   & 128 / 384   & Soft \\
                O$^{\ddag}$     &  \change{2021 June 71} & \change{59372.34} & 06:09-10:08 & 2.2 / 8.2   & 128 / 384   & Soft \\
                P     &  \change{2021 Nov 11} & \change{59530.90} & 19:38-23:37 & 4.9       & 512       & Hard \\ \hline
                \end{tabular}
                \\ \footnotesize{$^\dag$ \citep{Steiner2021J1803ATel,ShidatsuJ18032022}}
                \\ \footnotesize{$^{\ddag}$ Observations made using the dual S/X-band dichroic feed.}
            \end{table*}
            
        \subsubsection{ATCA}
            We observed J1803 using the Australia Telescope Compact Array (ATCA) on 18 dates during its 2021 outburst. Here, we present a subset of \changetwo{9 of} these observations between \change{2021 May 11} May and \change{2021 July 3} (MJD~59345--59398), taken during the rise phase and around the hard-to-soft X-ray state transition. During these observations, the ATCA was in the 750D, 1.5B, and 6B configurations\footnote{\url{https://www.narrabri.atnf.csiro.au/operations/array\_configurations/configurations.html}}. On all dates data were recorded at central frequencies of 5.5 and 9\,GHz, with a bandwidth of 2\,GHz in each frequency band. On 11th May we also observed at 16.7 and 21.2\,GHz. \changetwo{We observed in dual polarisation mode with orthogonal linear feeds, combining the XX and YY correlations to form Stokes I, which we focus on in this work.} The full data set will be presented in Espinasse et al. (in prep.).

            For all observations, we used PKS 1934-638 for bandpass and flux density calibration, and the nearby (5.9$\degree$ away) source B1817-254 for phase calibration. Flagging, calibration, and imaging were carried out following standard procedures with the Common Astronomy Software Application \citep[\textsc{casa}, version 5.1.3;][]{CASA2022}. We imaged using Briggs weighting with a robust parameter of 0 to balance sensitivity and resolution. To measure the flux density of the source, we fit a point source at the source position in the image plane using the \textsc{casa} task \texttt{IMFIT}, where we used the synthesised beam parameters as the elliptical Gaussian profile to fit to the source. We list this subset of measurements of J1803 in Appendix~\ref{sec:ATCA Data}

        \subsubsection{ALMA}
            J1803 was observed with the ALMA (Project Code: 2019.1.01324.T) on \change{2021 May 11} (epoch 1; MJD $59345.1951\pm0.0058$, 04:32:40--04:49:27 UTC) and \change{2021 May 15} (epoch 2; MJD $59349.2442\pm0.0056$, 05:43:33--05:59:51 UTC), for a total on-source observation time of 5.0 minutes per epoch. Data for both epochs were taken in Band 3, at a central frequency of 98.5\,GHz. The ALMA correlator was set up in the Frequency Division Mode (FDM) to yield $4\times2$ GHz wide base-bands, each with 1920$\times$0.976 MHz channels, and a 6.0-second correlator dump time. During our observations, the array was in its Cycle 7 C6 configuration, with 44/45 antennas on 2021 May 11/15. \changetwo{We observed in dual polarisation mode with orthogonal linear feeds. Given the low flux density of the source, the observations were not set up for polarisation calibration, and thus we were only able to combine XX and YY to form Stokes I.} We reduced and imaged (with natural weighting to maximize sensitivity) the data within the Common Astronomy Software Application package \citep[\textsc{casa} v6.2;][]{CASA2022}, using standard procedures outlined in the \textsc{casa}Guides for ALMA data reduction\footnote{\url{https://casaguides.nrao.edu/index.php/ALMAguides}}. We used J1924--2914 as a bandpass/flux calibrator, and J1752--2956 as a phase calibrator. Flux densities of the source were then measured by fitting a point source in the image plane (with the \texttt{imfit} task). MAXI J1803--298 was not detected on \change{2021 May 11}, with a $3\sigma$ upper limit of $1.5\,{\rm mJy\,bm}^{-1}$, but was clearly detected on \change{2021 May 15} with a flux density of $7.42\pm0.03\,{\rm mJy\,bm}^{-1}$. The non-detection and high RMS noise on 2021 May 11 was most likely due to the non-ideal weather conditions (high average precipitable water vapor of 4.2 mm on 2021 May 11, compared to 0.8 mm on 2021 May 15).
          
    \subsection{VLBA Imaging}
        We first imaged our VLBA data within \textsc{aips} using the CLEAN algorithm. On the days when there were multiple observations, we imaged each of these epochs separately, before concatenating them to increase sensitivity and $uv$-coverage, since they were separated by a short amount of time. We refer to these concatenated epochs by their first letter (e.g. epochs B1 and B2 became epoch B). J1803 is close ($\sim4\degree$) to the Galactic centre, and thus the longer baselines were affected by scattering due to the dense, turbulent interstellar medium, resulting in angular broadening. In order to recover images with a resolution that matched the effective resolution of the angularly broadened data, we applied a Gaussian $uv$-taper with 30\% power at 50 mega-wavelengths (the typical maximum baseline of these observations was 150--250 mega-wavelengths). While we tried many \textit{uv}-taper sizes, we chose this \textit{uv}-taper to maximize the recovered flux density in the images by suppressing the scattered long baselines, while not compromising the image quality by removing too many inner baselines. Thus, the resolution, as marked in the lower left corner of the final images (Fig.~\ref{fig:Combined Images}), is \change{lower} than in typical VLBA observations. We observed J1803 using the best fitting position from the \textit{Swift}/XRT localisation \citep{Gropp2021swfitJ1803}, although we shifted the phase centre of all of the observations to align with the centroid position of the compact component detected in epoch A. We were only able to reliably detect J1803 in epochs A, B, and C, likely the result of the transient jet ejecta adiabatically expanding and fading as they became resolved out. We henceforth focus on these three observations. We were unable to perform any self calibration to improve the significance of our detections, since the source was too faint in all of our VLBA epochs. 
        
    \subsection{Time-dependent Visibility Model Fitting}
        Following our basic characterisation of the images in \textsc{aips}, we successfully fit elliptical Gaussian model components directly to the complex visibilities in DIFMAP for each of the three VLBA epochs. The signal to noise was too low in each of these epochs to split the individual observations into time bins to track the intra-observational variability of the detected components. While model fitting is able to reduce the number of free parameters in the imaging problem relative to algorithms like CLEAN, this approach is still limited by having to perform model fitting on multiple individual time bins in order to detect motion and flux density variability. To overcome this issue, and to constrain the nature of the intra-observational variability in our VLBA data, we have developed a new model fitting approach in which we fit a single time-evolving model to the full set of visibilities from an observation.
        
        By parameterising the motion and flux density variability of modelled emission components, we are able to leverage information from all of the time bins in a single fit, rather than having to individually fit distinct models to each of the separate time bins. Since transient jets often appear in images as compact point sources or Gaussians, we are able to use these simple emission profile models, which have analytical representations in the visibility domain. We therefore fit these simple models directly to the time-stamped interferometric visibilities. We can allow any of the parameters that describe these model components (e.g. position, size, flux density) to be time-dependent, and explicitly parameterise their variability with analytic expressions that are included in the model. We can then predict the time-varying visibilities of our model and compare these to the measured visibilities of our observations, to fit these time-variable models directly to the underlying data without requiring any Fourier transforms. 
        
        In this work, we fit the position of the source components (modelled as circular Gaussians) with a ballistic (i.e constant velocity) model, in a polar coordinate system. While later-time deceleration of jet ejecta has been seen \citep[e.g.][]{Espinasse2020RelativisticJ1820+070}, on these scales, a constant velocity model is adequate. Our equations of motion for the position of the source at time $t$ were therefore,
        \begin{equation}\label{eqn:ballistic velocity model x}
            \Delta x(t) = \left[x_0 + \dot{r}\left(t-t_0\right)\sin\left(\theta\right)\right] \cos\delta,
        \end{equation}
        and
        \begin{equation}\label{eqn:ballistic velocity model y}
            \Delta y(t) = y_0 + \dot{r}\left(t-t_0\right)\cos\left(\theta\right),
        \end{equation}
        where $\Delta x(t)$ and $\Delta y(t)$ are the positions of the component, relative to the phase centre of the observations, in the directions of Right Ascension (RA) and Declination (Dec.), respectively; $x_0$ and $y_0$ are the fit positions of the source at the beginning of the observation, relative to the phase centre; $t_0$ is the time at the beginning of the observation; $\dot{r}$ is the proper motion of the fitted source, $\theta$ is the position angle on the sky along which the model component moves in \degree East of North; and $\delta$ is the declination of the source. The phase centre of the observations was shifted to align with the centroid position of the component detected in the image of epoch A. In our model, we allowed the flux density to vary linearly as,
        \begin{equation}\label{eqn:linear flux density model}
            F(t) = F_0 + \dot{F}(t-t_0),
        \end{equation}
        where $F_0$ is the flux density of the source at the beginning of each observation and $\dot{F}$ is its derivative. We can easily make this model more sophisticated by adding further degrees of freedom to account for variability in other model parameters, for example by allowing for expansion. However, in this work we only consider simple models that evolve linearly with time in position and flux density, due to the low signal to noise and particularly sparse \textit{uv}-coverage of our observations. Therefore our fit parameters for this model (and associated units) are $x_0$ (mas), $y_0$ (mas), $\dot{r}$ (mas\,\perhour), $\theta$ (deg.), $F_0$ (mJy), $\dot{F}$ (mJy\,\perhour), and the full-width half maximum angular size of the circular Gaussian component, $\theta_{\text{FWHM}}$ (mas). 
        
        In order to perform parameter estimation for these time-evolving models, we used the framework of Bayesian inference \citep[see][for a primer on Bayesian inference]{vandeSchoot2021BayesianPrimer}. Due to the phase referencing, we assumed that the data were well calibrated and thus were not corrupted by incorrect station-based complex gains, and instead only considered thermal noise. Since the thermal errors on complex visibilities are Gaussian \citep{Thompson2017InterferometryAstronomy}, we used a Gaussian likelihood in our parameter estimation. We describe our model formalism and application of Bayesian inference in the visibility domain in Appendix~\ref{sec:appendix math}. Station-based complex gains could be included in the model as fitted parameters, or could even be avoided completely by the use of calibration-independent closure quantities as the data products \citep[e.g.][]{Chael2018InterferometricAmplitudes}; however we do not consider that in this work. In order to explore the posterior probability distribution to estimate the best fitting model parameters, we used the Bayesian inference algorithm nested sampling \citep{Skilling2006NestedSampling} implemented in the \textsc{dynesty}\footnote{\url{https://github.com/joshspeagle/dynesty}} \citep{speagle2020Dynesty} Python package. Nested sampling is well suited for our model fitting requirements, given its ability to efficiently traverse multi-modal posterior distributions. We used the python library \textsc{eht-imaging}\footnote{\url{https://achael.github.io/eht-imaging/}} \citep{Chael2018InterferometricAmplitudes} for the handling of our UVFITS data and for pre-processing. Our implementation is available via GitHub\footnote{\url{https://github.com/Callan612/MAXIJ1803-Model-Fitting}}.

        \subsubsection{Validation with Synthetic Data}\label{sec:validation}
            Before applying this new approach to our VLBA observations of J1803, we first validated our technique with synthetic data sets. We generated these synthetic data within \texttt{eht-imaging}, by calculating model visibilities according to the $uv$-coverage of epoch A, and applying thermal noise. We used \texttt{eht-imaging} to generate synthetic observations with a broad range of model parameters to verify the validity of our approach. We included models with both elliptical and circular Gaussian components with fixed size and shape over the length of the observation. We also generated models in which the component was stationary, to confirm that with our radial parameterisation we would be able to identify stationary components. We also generated synthetic observations with model components that were variable in size.  We show the marginal and joint posterior probability distributions of the fits to two example synthetic data sets in Appendix~\ref{sec:appendix synthetic}. We found that in all of our simulations we were able to reliably recover the input model parameters with our model fitting, in both the high and low signal-to-noise regimes, including when components are stationary.
            
        \subsubsection{Application to VLBA Observations}
            Following our validation, we applied our technique to the VLBA observations of J1803. For all of our model fitting on J1803 we averaged the data to 60 seconds, to try to improve the signal to noise of the measured visibilities and to increase the speed of convergence. For both epochs A and B we fit a model consisting of a single Gaussian component with ballistic motion, and linearly varying flux density. While we tested models on synthetic data sets with both elliptical and circular Gaussians, we found that when working with the real data, we were unable to reliably constrain models consisting of an extended elliptical Gaussian component. This was likely due to the source being faint and the sparse $uv$-coverage, and so we chose to use circular Gaussian source models, reducing the number of free parameters. We similarly found that we were unable to reliably constrain any models that allowed for expansion of the Gaussian components, and thus we kept the size of the Gaussian components constant. 
            
            We experimented with a range of models for epoch C. We tried to fit a model similar to the models in epochs A and B, however we found that we were unable to constrain a reliable and informative solution. We also tried a range of models that included fixing the flux density, the position angle, and the size of the component in epoch C, however we were unable to constrain any reliable models that included motion of the component. We therefore fit a static circular Gaussian with linearly varying flux density.
            
            In epoch B, where we found evidence of motion of a jet component, we placed a weakly informative Gaussian prior on the position angle along which the component could move, based on our initial imaging. We also checked that the posterior distribution did not differ significantly when we used a uniform prior on the position angle. We placed Gaussian priors on the positions of the components based on their positions in the initial images. We placed uniform priors on all other parameters. We list the priors on all of our parameters in Table~\ref{tab:priors}. We show the marginal and joint posterior probability distributions of our fitted parameters in Appendix~\ref{sec:appendix real}. We report our best fit parameters as the median of the marginal posterior distribution, with the 1$\sigma$ uncertainties taken as the range between the median and the 16th/84th percentile.

    \subsection{X-ray Data}
        We also analyze available X-ray data on J1803 to track the evolution of the outburst, as an auxiliary source of information guiding the interpretation our VLBA observations. This includes publicly available light curves from MAXI/GSC\footnote{http://maxi.riken.jp/} \citep{Matsuoka2009MAXImission}, and \textit{Swift}/BAT\footnote{https://swift.gsfc.nasa.gov/results/transients/} \citep{Krimm2013BATtransient}, along with data from \textit{Swift}/XRT.
        
        We reduced \textit{Swift}/XRT data and extracted a light curve in the 0.3-10 keV band for J1803 using HEASOFT (v6.29) and the UK \textit{Swift} Science Data Centre online platform\footnote{https://www.swift.ac.uk/user\_objects/} \citep{Evans2007XRTlightcurve1, Evans2009XRTlightcurve2}. For extraction, we used the source coordinates from \citet{Gropp2021swfitJ1803} and binned the light curve by averaging per each individual observation. Hardness was calculated as the ratio between count rate in the 1.5-10 keV band over that in the 0.3-10 keV band.

\section{Results} \label{sec:results}
    \subsection{Validation Examples}
        We describe two of our synthetic observations to demonstrate the capability of our model fitting technique. We used model parameters that are similar to the parameters that we found best described our real observations of J1803. First we demonstrate the model fitting results for a circular Gaussian moving with a proper motion of 1~mas\,\perhour, along the $y$-axis ($\theta=0\degree$). Fig.~\ref{fig:synth moving corner} shows the joint and marginal posterior probability distributions for the fit to this synthetic data set with the true model parameters overlaid. In all but two parameters, the model fitting is able to recover the true parameter values within a $1\sigma$ credible interval, and in all of them within $2\sigma$. 
            
        We also demonstrate the model fitting results for an identical circular Gaussian, but with no motion. Fig.~\ref{fig:synth static corner} shows the joint and marginal posterior probability distributions for this fit. The key identifier for a lack of motion in the source is in the marginal posterior for $\dot{r}$, which is consistent with having a mode at the boundary value of zero. In the case where there is no motion, the position angle $\theta$ along which the component moves should be uniformly distributed. In the high signal-to-noise case we found this to be true, however there is a clear structure in the marginal posterior distribution for $\theta$ as seen in Fig.~\ref{fig:synth static corner}. This is likely the result of the sparse $uv$-coverage and poor signal to noise of the simulated observation. Motion on the plane of the sky manifests in the data as a change in the slope of the phase of the complex visibilities. When the signal is faint and the $uv$-coverage is sparse, as is the case with this synthetic observation, it is harder to constrain motion along the direction in which the $uv$-coverage is most sparse, resulting in the structure in the posterior distribution for $\theta$ seen in Fig.~\ref{fig:synth static corner}. This is analogous to imaging, where the shape and orientation of the resolution element of the image (which is described by the restoring beam) is determined by the distribution of the $uv$-coverage, which results in poorer resolution in the direction in which the $uv$-coverage is most sparse. Synthetic data sets generated with more complete $uv$-coverage and higher signal to noise showed a marginal probability distribution for $\theta$ that was much more uniform.

    \subsection{VLBA Images and Model Fitting}
        \begin{figure*}
            \centering
            \includegraphics[width=\linewidth]{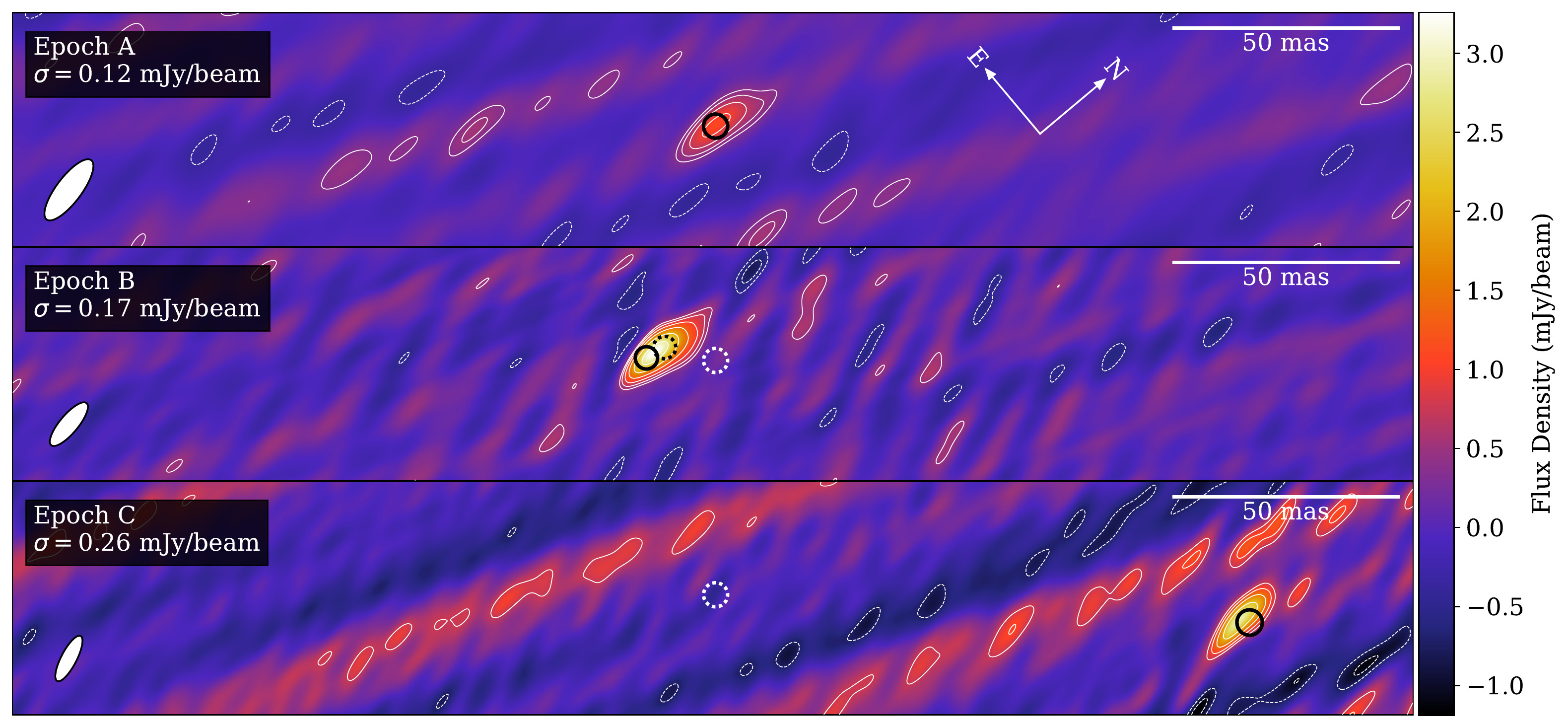}
            \caption{Images from epochs A, B, and C. Images have been rotated $50\degree$ clockwise (see arrows in the first panel), and all share the same colour scale as marked by the colour bar on the right. In the lower left corner of each image, a white ellipse marks the synthesised beam. The contours in each image are $\pm\sigma\times(\sqrt{2})^n$~mJy/beam for $n=3,4,5,6,...$, where $\sigma$ is the rms noise indicated in top left of each image. The black marker in the top panel shows the best fit location of a circular Gaussian in epoch A, which showed evidence that it was stationary. We plotted this same position with a white marker in the subsequent epochs. The dotted black marker in the second panel shows the best fit position of a moving circular Gaussian at the start of epoch B, and the solid black marker shows the position of the circular Gaussian at the end of the epoch. In the third panel we mark the position and size of the best fitting static circular Gaussian component with a black circle, since we were unable to constrain the motion of the component in this epoch. These plots do not capture the uncertainties associated with the model fits. The parameters of these model fits are listed in Table~\ref{tab:model fits}. There is evidence of motion of a jet component in epoch B, while our fits suggest the component in epoch A is the core of J1803. In epoch C we were unable to reliably constrain the motion of the component, however it does appear at a significant separation from the assumed core position, suggesting it is also a jet component. We discuss the identification and behaviour of these VLBA components in Sections~\ref{sec:identification} and \ref{sec:jet behaviour}.}
            \label{fig:Combined Images}
        \end{figure*}
        
        \begin{table*}
            \centering
            \caption{Best fitting model parameters. The reported value is the median of the marginal posterior distribution and uncertainties are the $1\sigma$ credible intervals. The positions $x_0$ and $y_0$ are given relative to the phase centre, which was shifted to centre on the centroid position of the component in the image of epoch A, at the coordinates, RA (J2000) $=18^\text{h} 03^\text{m} 2.79177^\text{s}$, and Dec. (J2000) $=-29\degree 49^\prime 49.41300^{\prime\prime}$.}
            \begin{tabular}{c|c|c|c|c|c|c|c}
                \hline
                Epoch & $F_0$                  & $\dot{F}$     & $x_0$                    & $y_0$                   & $\theta_{\text{FWHM}}$                    & $\dot{r}$              & $\theta$  \\
                      & (mJy)                  & (mJy\,\perhour)      & (mas)                    & (mas)                   & (mas)               & (mas\,\perhour)            & (\degree) \\ 
                \hline\hline
                A     & $0.76_{-0.18}^{+0.19}$ & $0.59\pm0.12$ & $0.19_{-0.5}^{+0.4}$     & $0.9_{-0.7}^{+0.8}$     & $5.2_{-0.4}^{+0.5}$ & $<0.48$ $(3\sigma)$ & -         \\
                B     & $5.1_{-0.2}^{+0.3}$    & $-0.02\pm0.1$  & $10.72\pm0.13$          & $-5.8\pm0.3$            & $4.4\pm0.2$         & $1.37\pm0.14$       & $168_{-4}^{+3}$ \\
                C     & $6.0\pm0.3$            & $-0.2\pm0.3$  & $-89.82_{-0.08}^{+0.09}$ & $84.59_{-0.15}^{+0.16}$ & $5.5\pm0.2$         & -                   & -         \\\hline             
            \end{tabular}
            
            \label{tab:model fits}
        \end{table*}
        
        We present, in Fig.~\ref{fig:Combined Images}, our CLEAN images and a visualisation of our model fitting results for epochs A, B, and C. In Table~\ref{tab:model fits} we detail the results of our model fits. We discuss the identification and behaviour of these VLBA components in Sections~\ref{sec:identification} and \ref{sec:jet behaviour}. The image of Epoch A consisted of a single compact source with a peak intensity of $1.2\pm0.1$~mJy/beam, where the uncertainty is the $1\sigma$ statistical uncertainty reported by the AIPS task JMFIT. In epoch A we allowed for a moving circular Gaussian, however we found that the marginal posterior probability distribution for $\dot{r}$ was consistent with having a mode at zero, with a $1\sigma$ credible upper limit on the motion of the component in epoch A being $\dot{r}<0.48$ mas\,\perhour. The marginal posterior probability distribution for $\theta$ (Fig.~\ref{fig:epoch A corner}) is not uniform and shows two peaks, the narrowest and largest at $\sim150\degree$ east of north and a small but broader peak at $\sim10\degree$ east of north. The posterior probability in between these peaks is non-negligible. The bi-modal peaks of this posterior distribution approximately correspond to the position angle of the CLEAN synthesised beam, i.e. the directions along which the $uv$-coverage is more sparse. This is identical to the behaviour seen in the synthetic observation described in Section~\ref{sec:validation}, suggesting that this component is most likely stationary. The best fit location of this component was,
        \begin{equation*}
            \text{RA (J2000)} = 18^\text{h} 03^\text{m} 2^\text{s}.79178\pm0.00003, 
        \end{equation*}
        \begin{equation*}
            \text{Dec. (J2000)} =-29\degree 49^\prime 49^{\prime\prime}.41220_{-0.00007}^{+0.00008}.
        \end{equation*}
        
        While we found the component in epoch A was most likely stationary, it was rapidly rising in flux density, with $F_0=0.76_{-0.18}^{+0.19}$~mJy and $\dot{F}=0.59\pm0.12$~mJy\,\perhour. The circular Gaussian component had a full-width half-maximum of $5.2_{-0.4}^{+0.5}$ mas. We tried splitting epoch A in half, and performed model fitting on each half independently, and found that we were able to consistently constrain the lack of motion and the rapid rise in flux density of the source in both halves of the observation. This epoch was the faintest of the three epochs, however it enjoyed the most sensitivity due to its observation duration. 
        
        In the image of epoch B, we found a single component $\sim10$ mas to the south-east (to the left in the rotated image) of the component in epoch A. It was also slightly extended with an asymmetric flux density distribution skewing towards the south-east. The peak intensity of the component was $4.0\pm0.2$~mJy/beam. In epoch B we found evidence of motion in our model fitting, with $\dot{r}=1.37\pm0.14$~mas\,\perhour\ at a position angle of $168\pm4\degree$ east of north. The position angle along which the components in epochs A, B, and C lie is approximately $135\degree$ east of north. The full-width half-maximum of the Gaussian component in epoch B was $4.4\pm0.2$ mas. Epoch B had the longest lever arm in time to detect motion, since in epoch A we observed geodetic blocks at the beginning and end of the observation. Epoch B, however, enjoyed less sensitivity than epoch A since it was comprised of two $\sim1$ hour long epochs separated by a $\sim1.5$ hour gap. 
        
        The image of Epoch C consisted of a single component $118$ mas away from the component in epoch A, to the north-west (right in the rotated image). Similar to epoch B, the component also appeared to be slightly resolved, although not as extended as the source in epoch B. It had a peak intensity of $3.8\pm0.3$~mJy/beam. Since we were unable to adequately converge on a solution for a moving circular Gaussian component, we fit a single static circular Gaussian with a full-width half-maximum of $5.5\pm0.2$ mas, at a separation of $122.0_{-1.7}^{+1.9}$~mas from the component in epoch A at a position angle of $313.9_{-0.6}^{+0.8}\degree$ east of north. As marked in the images, the best model fit position of the source is $\sim4$~mas further away from component A than the component in the image. It is unclear why this is the case, although it could be because of some faint, extended, asymmetric emission which we cannot detect in the imaging but can constrain in the model fitting. The image of epoch C contains some bright fringes with a direction and angular separation consistent with originating from the shortest baseline (LA-PT), possibly hinting at the existence of some larger extended emission that is only detected on the shortest baseline. The image of epoch D was dominated by this fringing, and we were unable to detect any compact source structure, which may be the result of the component in epoch C expanding and only being detected on this shortest baseline. In epoch C, the circular Gaussian component had a full-width half-maximum size of $5.5\pm0.2$ mas. The model fits for epoch B and epoch C were both consistent with having flat light curves, unlike epoch A where we saw a rapid rise in the flux density of the source over the length of the observation. Epoch C was the shortest and least sensitive of the three epochs in which J1803 was clearly detected, which could explain why we were unable to constrain a time-evolving model with motion.

        \change{We produced images for epochs A, B, C, and D, having flagged out the LA-PT baseline, which removed the large scale fringing. We found that for epoch C, the integrated flux density in the image was slightly reduced. We were still unable to detect any compact components in epoch D. Our modelling results were completely consistent with, and without the LA-PT baseline.}
        
        By stacking all of the observations of J1803 in the soft state \citep[MJD 59362--59500;][]{ShidatsuJ18032022}, we were able to put a 3$\sigma$ upper limit on the flux density of the core in the soft state of $<0.097$~mJy (at 8.2\,GHz). In the final observation, epoch P, we were unable to detect the core of the system in the hard state with a 3$\sigma$ upper limit of $0.105$~mJy (at 4.9\,GHz). 

    \subsection{Radio and X-ray Light Curves}
\begin{figure}
            \centering
            \includegraphics[width=\linewidth]{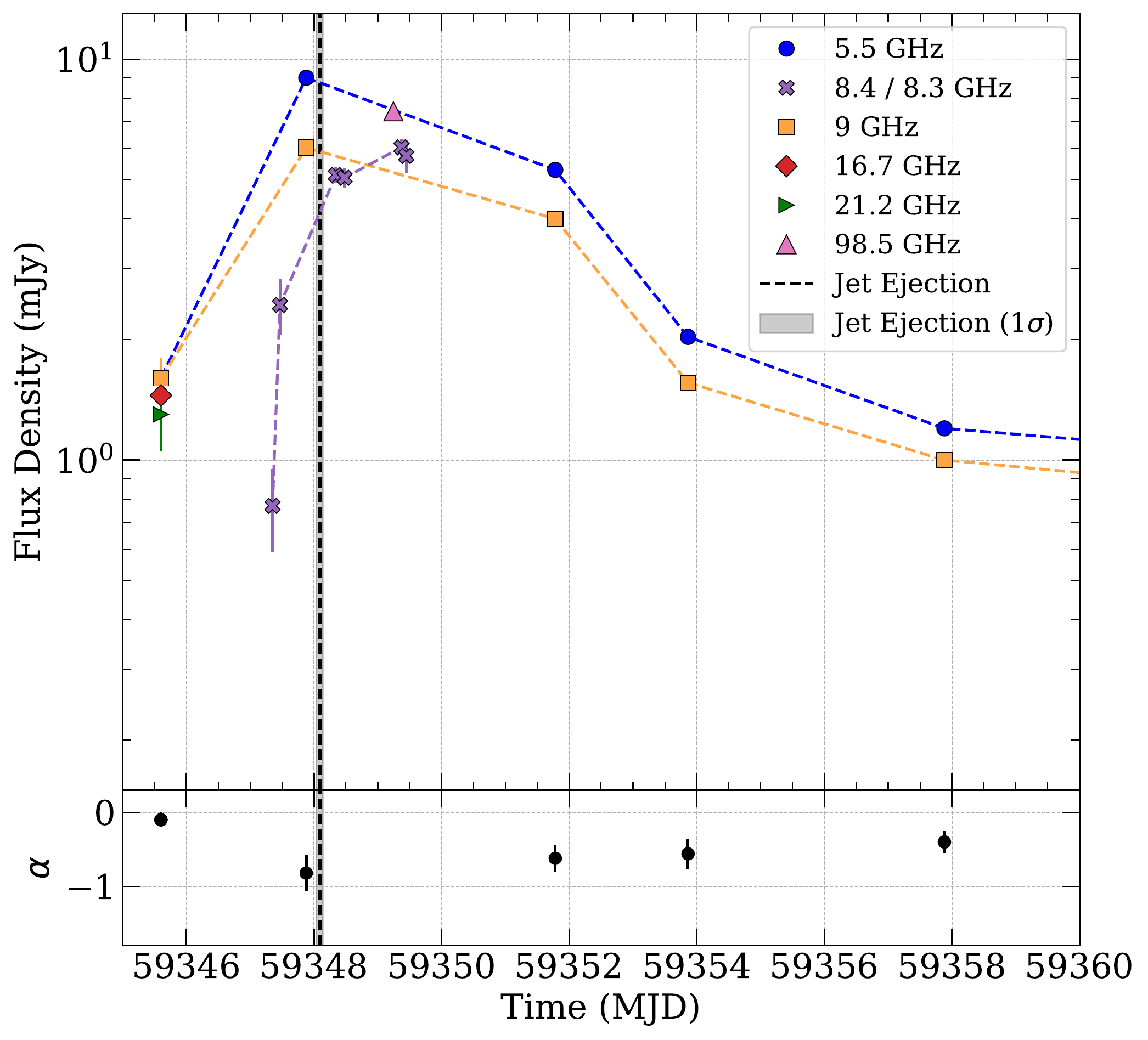}
            \caption{ATCA, ALMA, and VLBA light curves of the radio flare of \mj. The top panel contains the flux density measurements of our ATCA and VLBA observations. For the VLBA observations, we fit models where the flux density varied linearly, so we plot the flux density at the beginning and end of each observation. The bottom panel shows the spectral index calculated from the ATCA observations. The dashed vertical line and surrounding grey region mark the inferred ejection date of \change{the component in epoch B} (see Section~\ref{sec:jet behaviour}). The ejection occurs after the peak of the radio flare and the spectral steepening measured by the ATCA. Just prior to this peak, the VLBA light-curves show that the flare is rapidly rising.}
            \label{fig:Radio Flare}
        \end{figure}

\begin{figure}
            \centering
            \includegraphics[width=\linewidth]{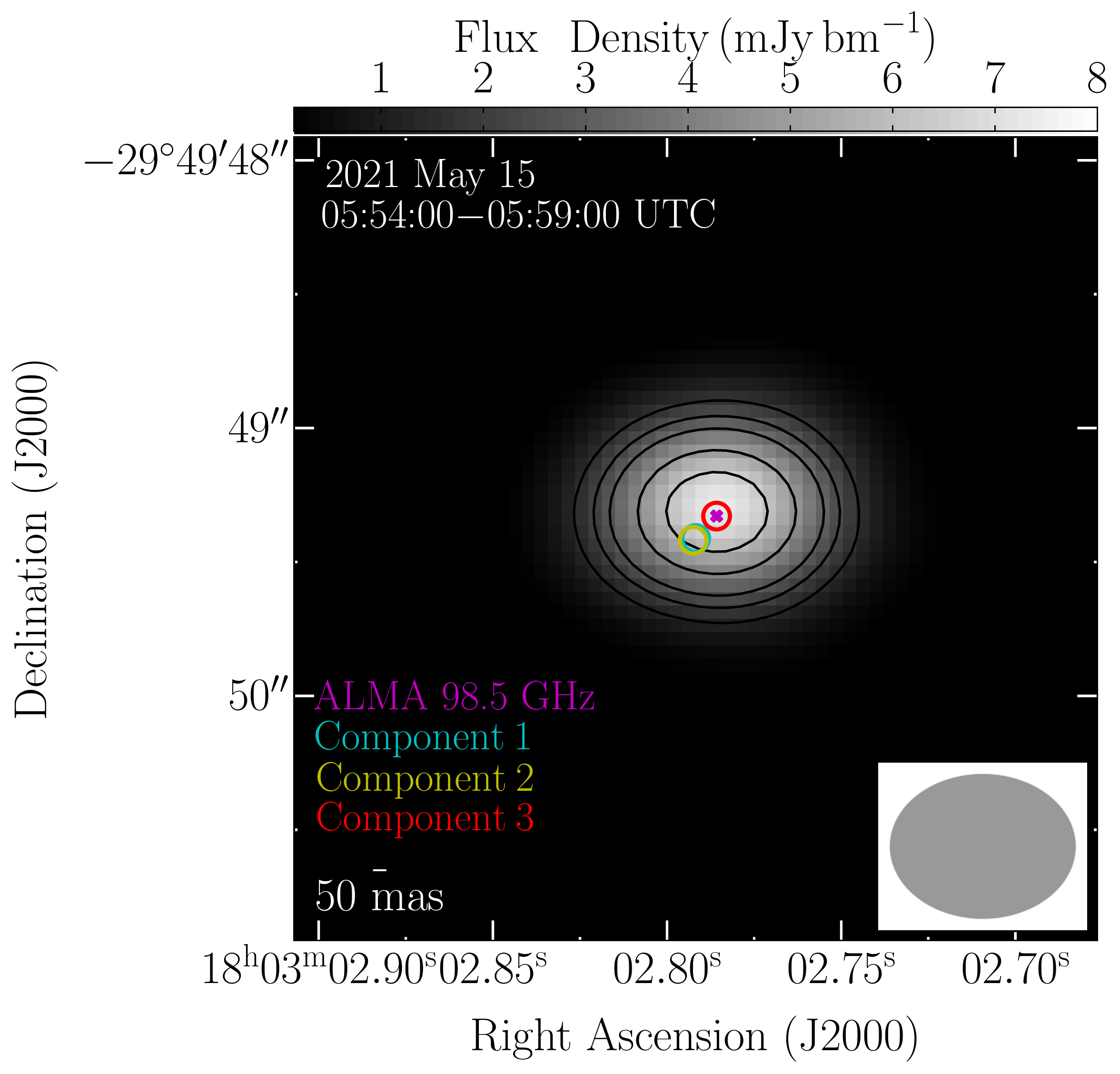}
            \caption{ALMA image of \mj\ from \change{2021 May 15} (MJD $59349.2442\pm0.0056$), observed at 98.5\,GHz. The ellipse in the lower right corner marks the synthesized beam. The positions of the components detected in the VLBA epochs A, B, and C, marked by the coloured circles and labelled components \change{1, 2, and 3}, respectively. The purple cross marks the centroid position of the point source component detected in the ALMA image. This position is consistent with the position of component \change{1}.}
            \label{fig:ALMA Epoch 2}
        \end{figure}

        In Fig.~\ref{fig:Radio Flare} we show the radio flare at the peak of the outburst, observed in both our ATCA and VLBA observations. Since we fit a linearly evolving flux density model to our VLBA observations, we plot the flux density at the beginning and end of each of the three VLBA epochs in which the source was detected. In the same figure we also show the evolution of the spectral index, $\alpha$ ($S_\nu\propto\nu^\alpha)$, over the ATCA observations. In the first ATCA observation at the beginning of the radio flare, the source spectrum was flat ($\alpha=-0.1\pm0.1$). In the second ATCA observation, J1803 reached its peak flux density at both 5.5 and 9 GHz, and had the steepest spectrum, with $\alpha=-0.8\pm0.2$. The source then began to fade, with the spectral index flattening gradually but remaining negative. In the first ATCA observation, there was no evidence of intra-observational variability over the length of the observation. In the second observation, which lasted for $\sim1.5$ hours, both the 5.5 and 9\,GHz flux densities were rising steadily, from 8.6/5.5 mJy  to 10.0/6.3 mJy (5.5/9\,GHz), suggesting that this observation did not correspond to the true peak of the radio flare. We also include the ALMA epoch 2, 98.5\, GHz flux density measurement, which occurred very close to the VLBA epoch C observation. In Fig.~\ref{fig:ALMA Epoch 2}, we show the image of the ALMA epoch 2, with the positions of the components from the VLBA epochs A, B, and C marked on top. The position of the unresolved component detected by the ALMA is consistent with the position of the component detected in the VLBA epoch C. We split the ALMA epoch 2 into 30 second time-bins, which revealed the source had an approximately constant flux density over the length of this observation, which was consistent with the flat flux density profile of the nearby VLBA epoch C. 
        
        The VLBA observations provided denser sampling of the flux density of J1803 around the peak of this flare than the ATCA observations, with epoch A showing rapid intra-observation brightening at 8.4 GHz, just prior to the peak of the ATCA radio flare. Our model fitting revealed no significant intra-observational flux density evolution in VLBA epochs B and C. J1803 was in the intermediate state for the entirety of the time-span shown in Fig.~\ref{fig:Radio Flare}.
         
\begin{figure*}
            \centering
            \includegraphics[width=\linewidth]{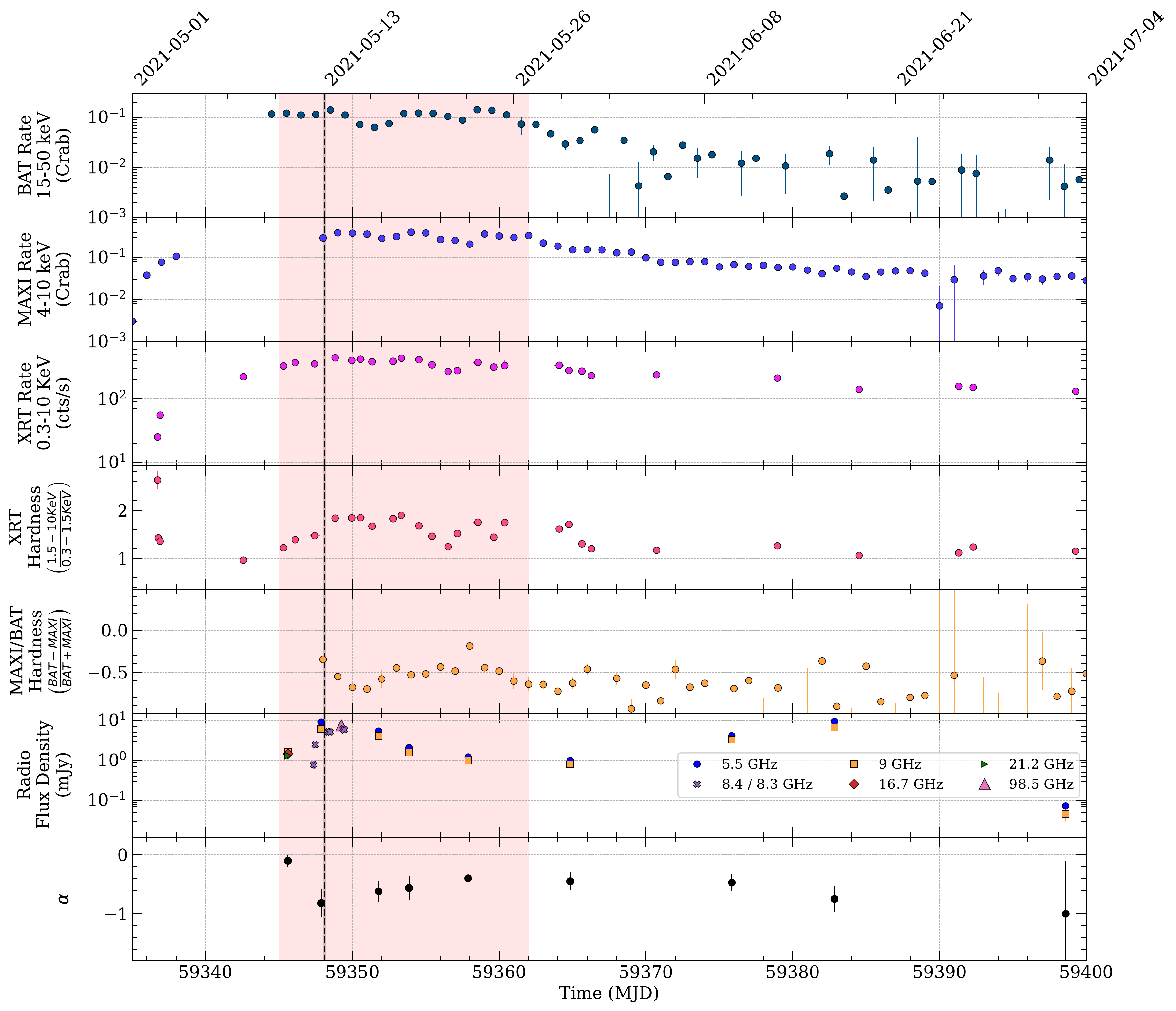}
            \caption{X-ray and Radio light curves surrounding the peak of the outburst of \mj\ and the subsequent state transition. The reverse transition is not included in the time range of this figure. The top three panels contain the light curves from the \textit{Swift}/BAT, MAXI/GSC, and \textit{Swift}/XRT telescopes, respectively \citep{ShidatsuJ18032022}. The next two panels show the XRT hardness ratio and the MAXI/BAT hardness ratio. The final two panels show the same ATCA and VLBA light curves and the associated spectral indexes as shown in Fig.~\ref{fig:Radio Flare}, extended to show the evolution of the ATCA light curves over the full state transition and the first few weeks of the soft state. The dashed vertical line marks the inferred ejection date of the \change{component in epoch B} (see Section~\ref{sec:jet behaviour}). The uncertainty is roughly the same as the width of the dashed line, and is marked by the thin grey region. The red shaded region shows the period during which \mj\ was in the intermediate state, between the initial hard state and the soft state. Following the initial radio flare and ejection of the jet, there is a long term radio re-brightening.}
            \label{fig:Xray and Radio LC}
        \end{figure*}

        In Fig.~\ref{fig:Xray and Radio LC} we show the full ATCA light curves with the ALMA and VLBA measurements, accompanied by the \textit{Swift}/BAT, \textit{Swift}/XRT, and MAXI/GSC 1-day averaged light curves of J1803 around the peak of the outburst and the state transition. In the ATCA light curves we observed a re-brightening following the initial radio flare, beginning at $\sim$ MJD $59365$ and peaking on MJD $59384.8$. During this period of re-brightening, we did not detect any emission from a compact core or transient radio jet in our VLBA observations.
        
\section{Discussion} \label{sec:discussion}
    We have developed and implemented a new model fitting technique in which we fit time-varying model components directly to the visibilities of an interferometric observation, where we explicitly parameterise the evolution of the components. We first verified this technique with synthetic observations, before applying it to three VLBA observations of J1803 during its 2021 outburst. We next discuss this new technique, the results of our modelling of the VLBA observations, and the interpretation of these results within the context of the overall outburst of J1803. This context includes our ATCA monitoring, which observed a radio flare around the peak of the outburst, and observations in the X-ray band, which probed the evolution of the inner accretion flow.
    
    \subsection{Visibility Model Fitting}\label{sec:visibility modelling discussion}
        Our model fitting approach extends the traditional model fitting implementations of software like DIFMAP \citep{Shepherd1994DIFMAP} or UVMULTIFIT \citep{Marti-Vidal2014UVMMULTIFIT} by parameterising the variability and motion of model components over a full observation. In this way we were able to leverage information from a full observation to constrain this variability. Model fitting approaches have captured the variability of jets launched by LMXBs, for example in V404 Cygni. However, \citet{MillerJones2019V404} performed their model fitting separately on each individual time bin. This was possible for V404 Cygni because all of the individual components were bright ($10^1$-$10^3$ mJy) point-like sources. Our new approach is similar in premise to the new time-resolved imaging techniques developed to create movies from EHT observations of the super-massive black holes M87$^*$ and Sgr A$^*$, which seek to leverage information from the full observation to enhance the quality of each time-binned image by enforcing or parameterising continuity between time bins \citep[][]{bouman2017reconstructing, Johnson2017DynamicalInterferometry, Arras2022VariableM87}. Rather than trying to perform a full pixel-by-pixel image reconstruction, as these techniques do, we parameterise our data with simple model components, which greatly reduces the number of free parameters in the reconstruction. The use of these simple models is physically motivated, since transient jets launched by LMXBs often appear as point sources or compact Gaussian components in VLBI images. We therefore do not require the more sophisticated image reconstructions that have been developed to account for the complex turbulent flows and asymmetry in objects such as M87$^*$ and Sgr A$^*$.
        
        We validated our technique using synthetic observations that were designed to replicate the typical source behaviour and variability we see in observations of jets from LMXBs. We found that we were able to reliably recover the input model parameters for a range of different synthetic data sets. 
    
        With our initial CLEAN imaging we were able to measure the locations, sizes and flux densities of discrete components in our observations. However we were unable to characterise the variability of these components, since the signal-to-noise of the observations was too low, and the $uv$-coverage was too sparse for time binning, even with standard model fitting within DIFMAP. With our new model fitting approach, we have been able to measure the proper motions of the components in epochs A and B by fitting a ballistic proper motion model, as well as measuring the flux density evolution of the components in epochs A, B, and C, by fitting a linear model. We were unable to fit more complicated models, including models with acceleration/deceleration, non-linear flux density evolution, elliptical Gaussian components, or expanding Gaussian components, due to the sparse $uv$-coverage and low signal-to-noise of the observations. In observations with more complete $uv$-coverage and higher signal-to-noise, we could in future extend our technique to include these more sophisticated models. For consecutive observations where the same components are detected multiple times, we could even use this technique to perform a single fit across multiple observations. In the case of J1803, the measured positions and proper motions (or lack thereof) for the different components in the three VLBA observations in which we have a detection suggest that they are distinct.
    
    \subsection{Identification of VLBI Components}\label{sec:identification}
        From our VLBI imaging and model fitting, we have identified three separate components in epochs A, B, and C, which we will refer to as components \change{1, 2, and 3}, respectively. Based on our model fitting and imaging we identify the three components as follows. 
        
        Our modelling suggested that component \change{1} is most likely stationary, and is rapidly rising in flux density. Given its apparent lack of motion, its compact structure, its presence towards the beginning of the outburst, and the fact that component \change{2} is approximately moving away from its position, we suggest that this component is likely the core of the system. It is not clear if the radio emission is originating from a compact, steady jet prior to it becoming quenched in subsequent epochs, or if the rapidly rising flux density originates from the rise of radio emission from slow-moving transient ejecta. While we suspect that component \change{1} is the core, we were unable to confirm the position of the core with any other observations. In epoch P, we were unable to detect the core of the system in the hard state, nor were we able to detect the core by stacking together all of the epochs in which there were non-detections. No other instrument that observed J1803 provides enough angular resolution to give an independent confirmation of the location of the core. Our only constraint on the position of the core of the system is therefore our measured position in epoch A, given in Section~\ref{sec:results}. 
        
        Component \change{2} is a jet moving away from the location of the core, which we believe to be component \change{1}. We note that the direction of the motion of component \change{2} does not point directly back to component \change{1}. It is not clear if this is due to some misalignment between the images, some directional bias in the modelling introduced by the non-uniform nature of the $uv$-coverage, or if component \change{1} is truly not the core. We also note that the markers plotted on top of the images in Fig.~\ref{fig:Combined Images} do not capture the uncertainty of the model fits. If component \change{1} is not the core, then our only constraint on its position is that it should be along the axis of the motion of component \change{2}, in the opposite direction. 
        
        Although we were unable to constrain the motion of component \change{3} (or lack thereof), we suggest it is most likely a jet, given its transient nature, and its position $\sim120$ mas away from component \change{1} approximately opposite from the direction in which component \change{2} moves. Our failure to constrain models that parameterised the motion of component \change{3} is likely the result of the low signal-to-noise ratio of the observation as well as the fact that the jet was starting to become resolved out during this epoch. 
    
    \subsection{Behaviour of the Jets}\label{sec:jet behaviour}
        
        \subsubsection{Ejection of Component \change{2}}\label{sec:B ejection}
            Assuming that the stationary component in epoch A marks the location of the core, we can track the motion of component \change{2} back to its origin at the core and infer its ejection date. Since the motion of component \change{2} does not point directly to the position of component \change{1}, we take the so called `ejection date' to be the time at which component \change{2} is closest to component \change{1}. We note that without knowledge of the core position (i.e., if component \change{1} is not the core), we cannot perform this analysis. Under this assumption, we therefore calculate an ejection date of MJD~$59348.08_{-0.06}^{+0.05}$, which was $6_{-1.2}^{+1.4}$ hours before the beginning of epoch B, confirming that this jet component was not present in epoch A. We mark this date in the light curves of Figs.~\ref{fig:Radio Flare} and \ref{fig:Xray and Radio LC}. This ejection date occurs just after the peak of the radio flare measured by ATCA, however since the ATCA observations were sparse, and since the flux density was steadily rising during the ATCA observation at the peak of the outburst, this date likely does not correspond to the true peak of the radio flare of J1803. The ATCA observation on this date showed the steepest spectral index, which is a signature of the presence of an expanded, optically thin radio jet. The rise phase in the radio flares seen at the state transition of LMXBs is usually attributed to the adiabatic expansion of an optically thick synchrotron emitting plasma cloud, with the radio flare peaking as the self-absorbed synchrotron turnover of the adiabatically expanding plasma clouds moves through the observing band and the jet becomes optically thin \citep[e.g.][]{Tetarenko2019TrackingOutburst, Fender2019SynchrotronHoles, Bright2020AnJ1820+070}. Another model for these flares is the shock-in-jet model \citep[e.g.][]{Jamil2010IShocks:Model, Malzac2014TheShocks}, in which jet material is accelerated by internal shocks as it collides downstream with previously ejected, slowing moving jet material. In these models the rise and peak of these radio flares usually lags the ejection of the jet material.
            
            An \textit{AstroSAT} observation on MJD 59345--59346 suggested that J1803 was already in the intermediate state \citep{Jana2021ATelAstroSATJ1803}, and spectral analysis of MAXI/GSC observations suggested that the transition from the intermediate state to the soft state occurred between MJD 59361 and MJD 59362 \citep{ShidatsuJ18032022}. This implies that the ejection of component \change{2} occurred while J1803 was in the intermediate state and not during the transition from the hard state to the intermediate state, or from the intermediate state to the soft state. The timing of this ejection is consistent with the current view of state transitions and the ejection of transient jets in LMXBs, where transient jets are launched during the intermediate state as the source transitions from the hard state to the soft state \citep[e.g.][]{Fender2004TowardsJets, Corbel2004StatesXTE1650, Fender2009JetsXrays, MillerJones2012DiscjetH1743322, Russell2019DiskJetJ1535571, Homan2020AJ1820+070, Wood2021Varying, Carotenuto2021J1348jets}. In the 1-day averaged X-ray light curves of J1803 (Fig.~\ref{fig:Xray and Radio LC}), we see an increase in the \textit{Swift}/XRT hardness ratio following the ejection of component \change{2}. The 15--50 keV \textit{Swift}/BAT light curve also shows a slight jump in intensity following the ejection of component \change{2}. This is followed by a multi-day decrease in intensity in the 15--50 keV \textit{Swift}/BAT light curve, while the 4--10 keV MAXI/GSC intensity remains steady. Given the data gaps and low statistics in the X-ray coverage around this inferred ejection date, it is difficult to identify any clear accretion signatures associated with the ejection of component \change{2}. 

            One such accretion signature, that has been associated with the ejection of transient jets, are low-frequency quasi-periodic oscillations (QPOs) seen in X-ray observations \citep[see][for a review of low frequency QPOs]{Ingram2019QPOReview}. At the beginning of an outburst, type-C QPOs are usually present, which eventually disappear and are often replaced by type-B QPOs, which are thought to be related to the ejection of transient jet material \citep{Fender2009JetsXrays, MillerJones2012DiscjetH1743322, Russell2019DiskJetJ1535571}. Recently, \citet{Wood2021Varying} showed that the ejection of transient jet material in MAXI J1820+070 occurred contemporaneously with the switch from type-C to type-B QPOs. 

            Observations with AstroSat on \change{2021 May 11} and \change{2021 May 12} (MJD 59345 and 59346) revealed the temporal and spectral properties of J1803 were evolving as the source transitioned from the hard-intermediate state to the soft-intermediate state \citep{Jana2022MNRASAstroSatJ1803-298}. Around MJD~59346.4, \citet{Jana2022MNRASAstroSatJ1803-298} observed a distinct change in the QPO frequency, which was followed by a decrease in the QPO strength at the end of the observation (MJD~59346.7). A type-C QPO was observed in J1803 by the Hard X-ray Modulation Telescope \citep[\textit{Insight-}HXMT][]{Zhang2020HXMTOverview} with the Medium Energy (ME) instrument (5-30~KeV) on MJD~59346.4, but not in the next observation on MJD~59346.6 (Yingchen Xu priv. commm.). This change in the QPO properties and eventual disappearance of the type-C QPO occurred approximately 36-40 hours prior to our inferred ejection date for component \change{2}. 
        
        \subsubsection{Jet Kinematics}

            We have identified component \change{2} as a jet and suggest that component \change{3} is most likely a jet travelling in approximately the opposite direction. It is not immediately clear if \change{2} and \change{3} are bipolar counterparts, or if they originate from two separate single-sided ejection events. Assuming a distance of 8 kpc to J1803 (based on its proximity to the Galactic centre), the apparent velocity of component \change{2} projected on the plane of the sky is $1.52\pm0.16$c, where $c$ is the speed of light. Relativistic jets can have apparently super-luminal motion as a result of projection effects of a sub-luminal source moving towards us with sufficient speed and inclination \citep[e.g.][]{MirabelandRodriguez1994AGalaxy, Bright2020AnJ1820+070,Carotenuto2021J1348jets}. 
            
            If the proper motions of intrinsically symmetric bipolar relativistic jets can be measured, then the intrinsic velocity and inclination angle can be calculated \citep{MirabelandRodriguez1994AGalaxy}. If the distance to the source is not known, an upper limit on the distance can be computed by assuming a jet speed of $\beta=1$ and an inclination of $90\degree$, yielding,
            \begin{equation}
                d_\text{max} = \frac{c}{\sqrt{\mu_{\text{app}}\mu_{\text{rec}}}},
            \end{equation}
            \citep{Fender2003UsesMicroquasars}. In the case of J1803, the distance to the source is unknown, and we were only able to measure the motion of component \change{2}. However, since we were able to determine the ejection time of component \change{2} (under the assumption that component \change{1} was the core), we can try assuming that component \change{3} is the bipolar counterpart to component \change{2}, and thus was ejected at the same time, giving a proper motion of $3.96_{-0.17}^{+0.16}$ mas\,\perhour. This proper motion is larger than the proper motion of component \change{2}, which would make it an approaching counterpart to a receding component \change{2}. We then calculate a value of $d_\text{max}=3.10\pm0.17$ kpc. Again, we note that if component \change{1} is not the core of the system, we are unable to put any constraints on the motion of component \change{3}. Given the proximity of J1803 to the Galactic centre, the fact that it is scattered by the dense ISM, and that it is reasonably faint, this distance upper limit seems to be too low, which suggests that component \change{3} is likely not an approaching counterpart to a receding component \change{2}, and may not have been ejected at the same time as component \change{2}. For this distance upper limit to be closer to the expected 8\,kpc, assuming components \change{2} and \change{3} are symmetric counterparts, the proper motions of components \change{2} and \change{3} must be lower, and thus they must have been ejected much earlier, which could invalidate the assumption that component \change{1} was the core. If these components were ejected earlier they would still need to be consistent with being ejected during the intermediate state, since it is unlikely that the transient jets were launched during the hard state. We note that J1803 is located close to the centre of Baade's Window \citep{Baade1946Window} and thus the presence of a bright (15.82 magnitude) optical counterpart detected by \textit{Swift}/UVOT \citep{Gropp2021swfitJ1803} does not necessarily suggest that J1803 cannot be located close to the Galactic centre. 
        
            Based on the inconsistency of the calculated upper distance to J1803 with components \change{2} and \change{3} as symmetric counterparts to the assumed distance close to the Galactic centre, as well as the misalignment of the motion of component \change{2} with the position angle of the jet axis drawn between components \change{1} and \change{3} by $33_{-4}^{+3}\degree$, components \change{2} and \change{3} are unlikely to be symmetric bipolar counterparts, and are instead more likely to be single-sided ejecta. Without a symmetric counterpart for either components \change{2} or \change{3}, it is impossible to conclusively say whether or not either are approaching us or receding from us. Since relativistic jets will be Doppler boosted, if we only see a single jet component then it is likely approaching us. This is difficult to explain in these observations since \change{2} and \change{3} appear on opposite sides of the core, meaning they cannot both be approaching unless there is a large-scale Lense-Thirring precession of the jet axis. There is some evidence of precession of the jet axis, since the model fit for component \change{2} suggested that its position angle was misaligned with the position angle of the jet axis drawn between components \change{1} and \change{3} by $33_{-4}^{+3}\degree$. Precession of the jet axis has been seen before, such as in V404 Cygni \citep{MillerJones2019V404}, SS 433  \citep{Hjellming1981SS433}, and GRO 1655-40 \citep{Hjellming1995Episodic40}, with precession cone half opening angles of $\sim18\degree$,$\sim20\degree$, and $\sim2\degree$, respectively. 
            
            If there is no large scale precession of the jet axis, and components \change{2} and C are not bipolar counterparts, this may hint to some intrinsic asymmetry in the system. Although it is often assumed that jet ejections are symmetric, this assumption may not be generally true \citep{Fendt2013Asymmetricjets}. This could be due to an intrinsic asymmetry in the inner accretion flow and jet launching mechanism, or an asymmetry in the surrounding ISM with which the jet material interacts. One-sided jets have been observed before, such as in V404 Cygni \citep{MillerJones2019V404}, but just as in this case, it is not clear if this was the result of Doppler boosting, asymmetry in the surrounding ISM, or an intrinsic asymmetry of the jets themselves.
            
            As the jets moved away from the core they expanded, eventually becoming resolved out. In epoch C, we saw that there was some extended emission that was only detected on the shortest baseline, likely the result of the expansion of the jet component. Comparing the VLBA flux density of component \change{3} with the interpolated ATCA flux density at 9 GHz in Fig.~\ref{fig:Radio Flare} suggests that the VLBA measurement in epoch C was consistent with or even above the interpolated ATCA 9GHz flux density. We applied the VLBA epoch C flux density decay rate ($\dot{F}$) to the ALMA epoch 2 flux density measurements (both directly and by scaling the timescale according to the van der Laan plasmon model \citep{1966Natur.211.1131V} which causes the higher frequencies to decay quicker) to the predict the ALMA flux density at the time of the VLBA epoch C observation. This revealed an approximately flat spectral index at the time of epoch C, which indicates that we likely captured close to the peak of the flux density of the component in our VLBA images. 
            
            Figure~\ref{fig:Combined Images} shows a small discrepancy between the position of component \change{3} from the imaging and the modelling. We suggest that this is likely caused by a faint, extended, and asymmetric region of flux density that is resolved out in the imaging but is detected by our modelling. Another possibility is that the phase calibration in this epoch was poor due to the low elevation of the source. This would most affect the long baselines, which were suppressed in our imaging by the $uv$-taper, but were not excluded from our modelling. By epoch D, the jet seen in epoch C had expanded and become almost completely resolved out, with only a small amount of emission being detected by the innermost baseline. We were unable to detect jets in any subsequent epochs, despite the ATCA light curves showing that the system was well above our VLBA detection threshold for the entire duration of the May/June monitoring campaign. This suggests that the emission was resolved out in our VLBA observations due to this expansion. While we know that the jets were expanding, we were unable to fit models that explicitly parameterised this expansion within each of the VLBA observations, likely due to the low signal-to-noise ratio and sparse $uv$-coverage of the observations. 

            \change{In our modelling, we found that all three components were resolved with a significance of $>10\sigma$, which is consistent with our findings that J1803 was resolved in the images, due to interstellar scattering along the line of sight.}
            
            While we fit the full-width half-maximum size of the circular Gaussian components, it is difficult to use these to reliably measure the expansion speed and opening angle of the jets. With only a single size measurement for each of the three distinct components, we cannot disentangle their intrinsic size from the scattering kernel that is responsible for the angular broadening of the components. We also found that component \change{2} was smaller than component \change{1}, which may suggest in-homogeneity in the scattering screen, since if \change{component 1} is the core, it should be more compact than the jet, component \change{2}. We can, however, use the size and separation of components \change{2} and \change{3} from the core, to place upper limits on the jet opening angle, $\theta_j$, of $\theta_j<20\degree$ and $\theta_j<2.5\degree$ with components \change{2} and \change{3} respectively. This upper limit from component \change{3} is unreliable, since we suspect that there is further extended emission that is resolved out by the VLBA. 
            
            Jet opening angles have only been constrained for a number of black hole X-ray binaries \citep[see e.g.][]{stirling2001jets, Miller-Jones2006OpeningJets, Rushton2017ResolvedExpanding,  Tetarenko2017ExtremeCygni, Tetarenko2019CompactCygX-1, Espinasse2020RelativisticJ1820+070, Chauhan2021broadbandjet, Tetarenko2021MJ1820Timing}. Measurements of jet opening angles range from 0.4$-$1.8\degree in the compact jet of Cygnus X-1 \citep{Tetarenko2019CompactCygX-1} to $\sim58\degree$ in XTE J1908+094 \citep{Rushton2017ResolvedExpanding}, although we note that \citet{Rushton2017ResolvedExpanding} suggested that this number is derived from the lateral expansion of the jet lobe and thus the jet opening angle is likely smaller. Our constraints on the jet opening angle of J1803 are in agreement with the typical range of measurements of the jet opening angle for other black hole X-ray binaries.  
            
            Following the transition from the intermediate state to the soft state \citep[$\sim$ MJD 59362;][]{ShidatsuJ18032022}, the ATCA light curve showed a gradual rebrightening. Our VLBA monitoring continued through to the beginning of the rise phase of this secondary flare, and our lack of a detection of any newly ejected jet material leads us to conclude that this re-brightening was likely the result of one of the by then expanded, resolved out jet components interacting with the inhomogenous ISM downstream \citep[e.g.][]{Russell2019DiskJetJ1535571, Espinasse2020RelativisticJ1820+070, Bright2020AnJ1820+070, Carotenuto2022J1348deceleration}.

    \subsection{Behaviour of the Core}
        Since we do not detect the core in epoch B, and assuming that component \change{1} is the core of J1803, we can constrain the time period during which the core switched off to be between MJD~59347.49 and MJD~59348.34. We constrained the ejection date of component \change{2} to be within this period of time, however it is not clear if the quenching of the compact jet in the core coincided with the ejection of component \change{2}. In H1743-322, \citet{MillerJones2012DiscjetH1743322} observed that the quenching of the compact radio jet emission occurred during or immediately after the ejection of transient ejecta. In J1803, following the ejection of component \change{2}, the compact radio core was quenched for the remainder of the intermediate state and into the soft state. We were unable to detect the reestablishment of the radio core in our VLBA observations. 
        
        In epoch A, we observed the core to be rapidly rising in flux density at the beginning of the radio flare prior to the ejection of component \change{2}. As discussed in Section~\ref{sec:B ejection}, the rise of these radio flares is often attributed to either the expansion of a jet or the acceleration of jet material as it moves away from the launch site or the formation of internal shocks, and thus the rise generally is expected to lag the ejection of the jet material. The rapid rise in epoch A could be due to a sudden brightening of the compact jet prior to the ejection of component \change{2}, the adiabatic expansion of a much slower moving optically thick jet component close to the core, or internal shocks in jet material deposited close to the core of J1803. If component \change{1} is not the core, its rapid rise could be attributed to the evolution of an isolated, slow moving jet component, which may be seen later as either component \change{2} or \change{3}. We note that this component would have to undergo some acceleration prior to epoch B for this to correspond to component \change{2}. We cannot rule out that this may be a slow-moving component \change{3} present in epoch A, since we have no constraints on the motion of component \change{3}. 

\section{Conclusions}\label{sec:conclusion}
    We have developed a new model fitting approach by which we fit a continuous time-evolving model directly to the visibilities of an interferometric observation to parameterise and constrain intra-observational variability. We validated our technique with synthetic observations before applying it to three VLBA observations of the candidate black hole LMXB \mj\ during its May 2021 outburst. 
    
    With traditional imaging techniques we were able to locate a single discrete component within each epoch. However, due to the sparse $uv$-coverage and low signal to noise of these observations we were unable to perform time binning to capture any intra-observational variability. With our new model fitting approach, we were able to constrain the flux density variability of all three components, and constrain the proper motions of the components within the first two epochs. Based on our model fits we suggest that the component in the first epoch is a rapidly rising but stationary core, and the components in the second and third epochs are two distinct transient jets. We measured the proper motion of the jet in the second epoch to be $1.37\pm0.14$~mas\,\perhour. Based on the assumption that the component seen in the first epoch is the core of the system, we were able to infer the ejection date of the jet component from the second epoch to be MJD~$59348.08_{-0.06}^{+0.05}$. This occurred during the peak of the outburst, while \mj\ was in the intermediate state. 
    
    New VLBI imaging and model fitting techniques, like the one described here, are key for obtaining better constraints on the motions and inferred ejection dates of transient jets launched by LMXBs. With more precise ejection dates, coupled with dense, simultaneous X-ray monitoring and multi-wavelength radio coverage, we will be able to more clearly identify signatures of jet ejection, which will allow us to better understand the nature of the causal connection between changes in the inner accretion flow and the ejection of relativistic jets in black holes.

\section*{Acknowledgements}
The National Radio Astronomy Observatory is a facility of the National Science Foundation operated under cooperative agreement by Associated Universities, Inc. This work made use of the Swinburne University of Technology software correlator, developed as part of the Australian Major National Research Facilities Programme and operated under licence. The ATCA is part of the Australia Telescope National Facility, which is funded by the Australian Government for operation as a National Facility managed by CSIRO. This paper makes use of the following ALMA data: ADS/JAO.ALMA\#2019.1.01324.T. ALMA is a partnership of ESO (representing its member states), NSF (USA) and NINS (Japan), together with NRC (Canada), MOST and ASIAA (Taiwan), and KASI (Republic of Korea), in cooperation with the Republic of Chile. The Joint ALMA Observatory is operated by ESO, AUI/NRAO and NAOJ. This work made use of data supplied by the UK \textit{Swift} Science Data Centre at the University of Leicester. This research has made use of the MAXI data provided by RIKEN, JAXA and the MAXI team. 
CMW acknowledges financial support from the Forrest Research Foundation Scholarship, the Jean-Pierre Macquart Scholarship, and the Australian Government Research Training Program Scholarship. TDR acknowledges financial contribution from the agreement ASI-INAF n.2017-14-H.0. AJT acknowledges support for this work was provided by NASA through the NASA Hubble Fellowship grant \#HST--HF2--51494.001 awarded by the Space Telescope Science Institute, which is operated by the Association of Universities for Research in Astronomy, Inc., for NASA, under contract NAS5–26555. TMB acknowledges financial contribution from grant PRIN INAF 2019 n.15. FC acknowledges support from the Royal Society through the Newton International Fellowship programme (NIF/R1/211296). RS acknowledges grant number 12073029 from the National Natural Science Foundation of China (NSFC). VT acknowledges support from the Romanian Ministry of Research, Innovation and Digitalization through the Romanian National Core Program LAPLAS VII – contract no. 30N/2023.
The authors wish to recognize and acknowledge the very significant cultural role and reverence that the summit of Maunakea has always had within the indigenous Hawaiian community. We are most fortunate to have the opportunity to conduct observations from this mountain. We also acknowledge the Gomeroi people as the traditional custodians of the ATCA observatory site. 

%%%%%%%%%%%%%%%%%%%%%%%%%%%%%%%%%%%%%%%%%%%%%%%%%%
\section*{Data Availability}
The data used in this work are available via reasonable request to the corresponding author. Our implementation of the time-dependent model fitting is available via GitHub\footnote{\url{https://github.com/Callan612/MAXIJ1803-Model-Fitting}}. 



\bibliographystyle{mnras}
\bibliography{references} % if your bibtex file is called example.bib

%\begin{thebibliography}{99}
%\end{thebibliography}

%%%%%%%%%%%%%%%%%%%%%%%%%%%%%%%%%%%%%%%%%%%%%%%%%%

%%%%%%%%%%%%%%%%% APPENDICES %%%%%%%%%%%%%%%%%%%%%

\appendix

\section{Model Fitting Formalism}\label{sec:appendix math}
    We perform our parameter estimation by computing the posterior probability of each model, $P(\boldsymbol{\Theta}|V)$, using Bayes' theorem,
    \begin{equation}\label{eq:Bayes}
        P(\boldsymbol{\Theta}|V) = \frac{P(V|\boldsymbol{\Theta})P(\boldsymbol{\Theta})}{P(V)},
    \end{equation}
    where $\boldsymbol{\Theta}=\{\Theta_1,\Theta_2,...,\Theta_n\}$ is a vector of $n$ model parameters, $V$ is the array of measured visibilities, 
    \begin{equation}
        P(\boldsymbol{\Theta})=\prod_{k=1}^nP(\Theta_k)
    \end{equation}
    is the prior, and 
    \begin{equation}
        P(V)=\int P(V|\boldsymbol{\Theta})P(\boldsymbol{\Theta})d\boldsymbol{\Theta}
    \end{equation}
    is the evidence. The nested sampling algorithm \citep[][]{Skilling2006NestedSampling} is designed to efficiently approximate and compute the evidence integral, and as a by-product it returns a set of samples from the posterior distribution. Since the thermal noise on the visibilities is Gaussian \citep{Thompson2017InterferometryAstronomy}, and assuming that the measurements are uncorrelated, the likelihood is the product,
    \begin{equation}
        P(V|\boldsymbol{\Theta}) = \prod_t \prod_{ij}  \frac{1}{\sqrt{2\pi\sigma_{ij,t}^2}} \exp\left(-\frac{1}{2} \frac{\|V_{ij,t}^\prime(\boldsymbol{\Theta}) - V_{ij,t}\|^2}{\sigma_{ij,t}^2}\right)
    \end{equation}
    where $V_{ij,t}$ is the measured visibility from the telescope pair $ij$, at timestamp $t$, $\sigma_{ij,t}$ is the standard deviation of the thermal noise on that measured visibility, and $V_{ijt}^\prime(\boldsymbol{\Theta})$ is the model visibility generated for that telescope pair at that timestamp according to the model parameters $\boldsymbol{\Theta}$. Since the visibilities are complex quantities, we use the Euclidean $\|.\|_2$ norm. 
    
    In our model fitting for epochs A and B, we used a circular Gaussian source model, where the position of the component (relative to the phase centre), as a function of time, is given by equations \ref{eqn:ballistic velocity model x} and \ref{eqn:ballistic velocity model y}, and the flux density of the component, as a function of time, is governed by equation \ref{eqn:linear flux density model}. For epoch C, we kept the position of the component constant with time. Since a circular Gaussian in the image plane is also a circular Gaussian in the Fourier plane, we can analytically compute the model visibilities as a function of the component position, $\Delta x(t)$ and $\Delta y(t)$, the flux density, $F(t)$, and the full-width half-maximum size, $\theta_{\text{FWHM}}$, for each telescope pair $ij$, at time $t$. This gives,
    \begin{multline}
        V_{ij,t}^\prime = F(t) \cdot \exp\left(\frac{-\pi^2}{4\ln(2)} \theta_\text{FWHM}^2 \left[u_{ij,t}^2+v_{ij,t}^2\right]\right)\cdot \\
        \exp\left(2\pi\mathrm{i}\left[ u_{ij,t}x(t) + v_{ij,t}y(t)\right] \right)
    \end{multline}
    where $u_{ij,t}$ and $v_{ij,t}$ are the vector coordinates of the projected baseline of the telescope pair $ij$ at time $t$. This expression could be replaced by any given source model, for example a point source or an elliptical Gaussian.

\section{Synthetic Observation Posterior Distributions}\label{sec:appendix synthetic}
\begin{figure*}
    \centering
    \includegraphics[width=\linewidth]{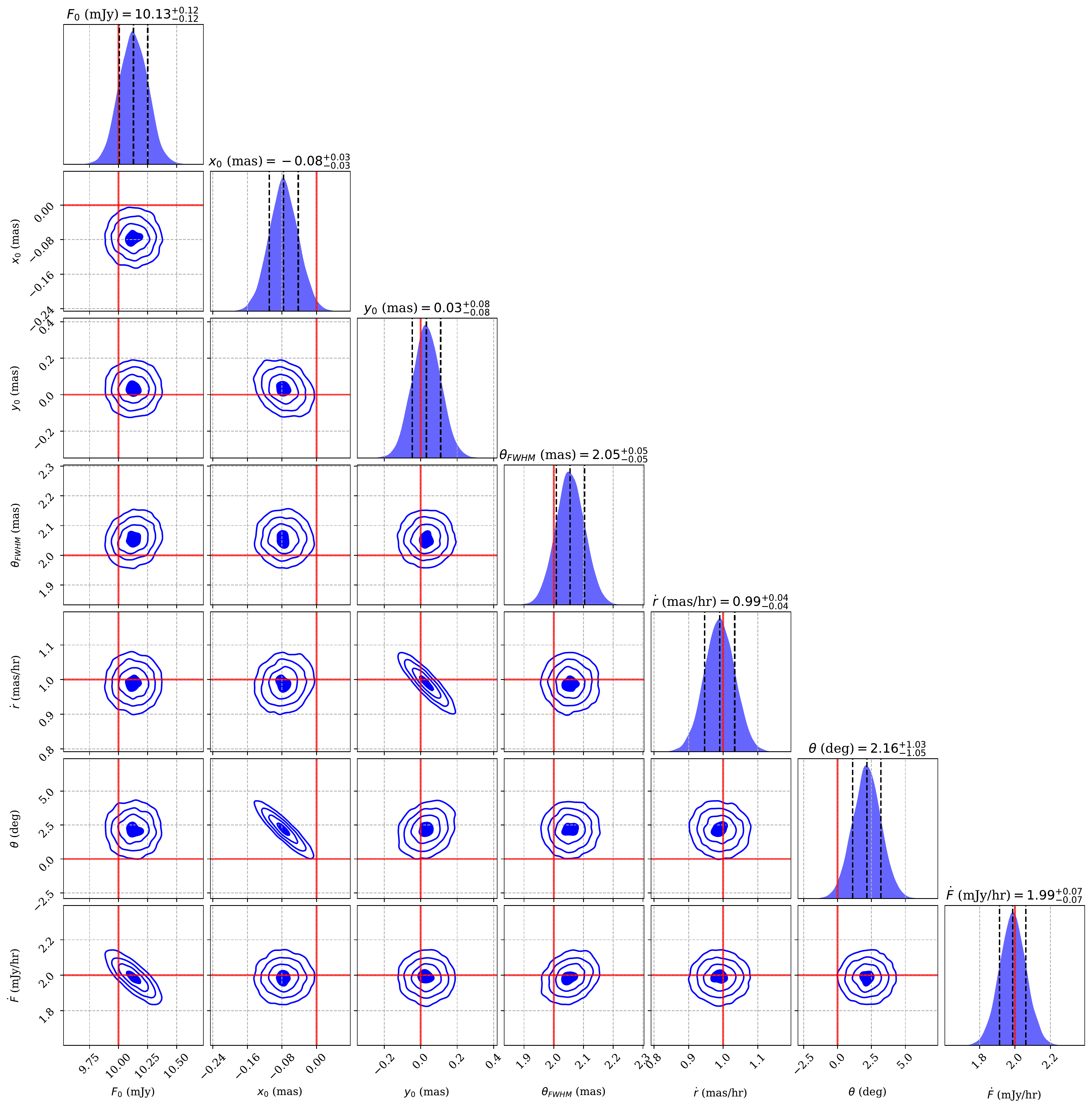}
    \caption{Marginal and joint posterior distributions for the fitted parameters for a synthetic observation of a moving source. The blacked dashed vertical lines mark the 16th, 50th, and 84th quantiles. The blue contours in the joint posterior distributions mark the 0.5, 1, 1.5 and 2$\sigma$ levels. The red vertical lines mark the true parameter values used to generate the observation. Our model fitting code is able to recover the input parameters for a synthetic observation with a moving source.}
    \label{fig:synth moving corner}
\end{figure*}

\begin{figure*}
    \centering
    \includegraphics[width=\linewidth]{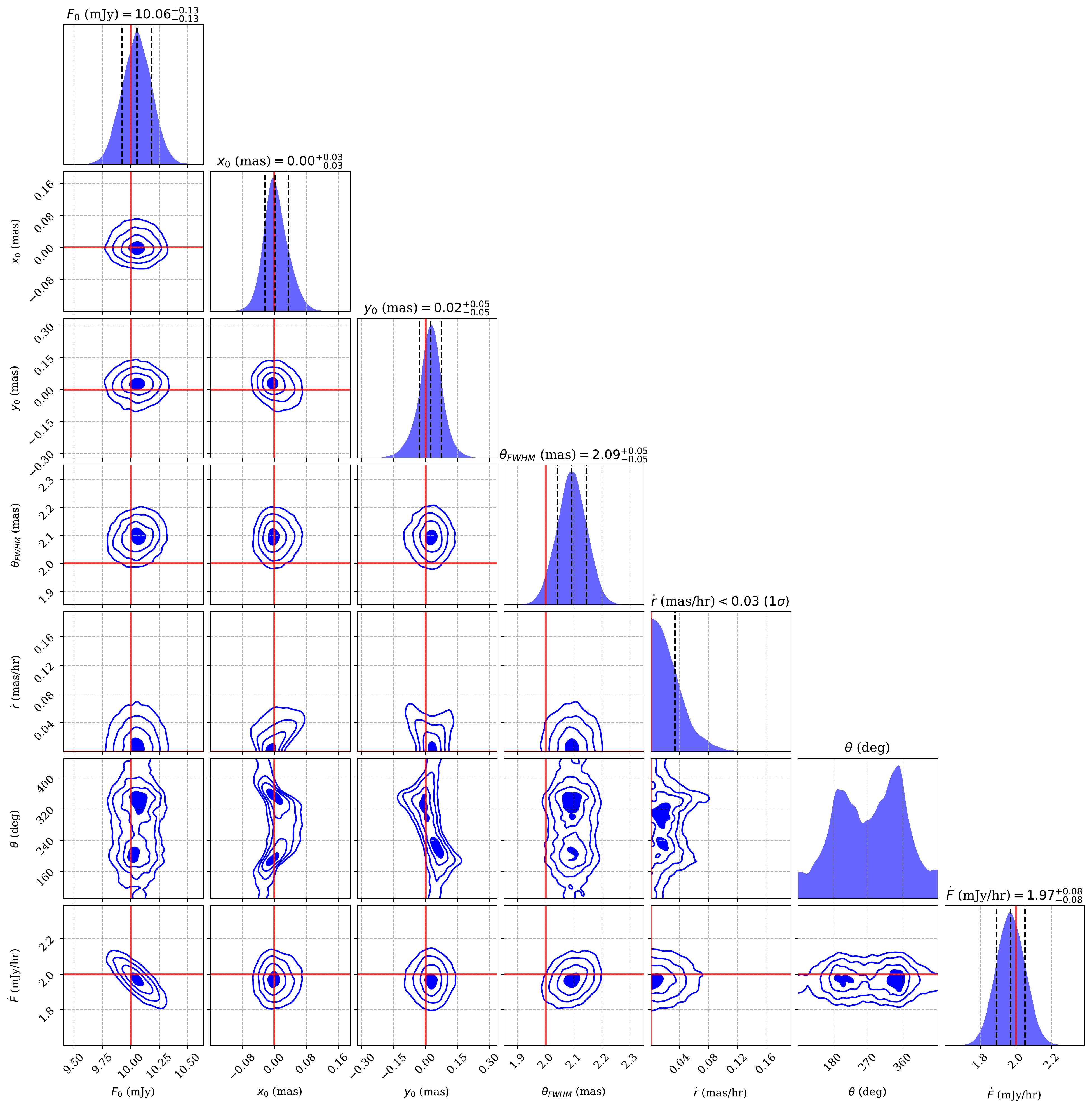}
    \caption{Marginal and joint posterior distributions for the fitted parameters for a synthetic observation of a stationary source. For $\dot{r}$, we mark the 1$\sigma$ upper limit, and for $\theta$ we mark no credible intervals. For all other parameters the black vertical lines mark the 16th, 50th, and 84th quantiles. The blue contours in the joint posterior distributions mark the 0.5, 1, 1.5 and 2$\sigma$ levels. The red vertical lines mark the true parameter values used to generate the observation. For a static source, the marginal posterior distribution for $\dot{r}$ is consistent with having a mode at 0. The joint probability distributions with $\theta$ have a unique non-Gaussian structure, as a result of the sparse $uv$-coverage used to generate the data.}
    \label{fig:synth static corner}
\end{figure*}

In Figs.~\ref{fig:synth moving corner} and \ref{fig:synth static corner}, we present the joint and marginal posterior distributions for our simulated observations of both a moving and static source, respectively, with the truth values overlaid. These simulated data included only thermal noise and were generated using the \textit{uv}-coverage of epoch A. Fig.~\ref{fig:synth moving corner} shows that we are able to recover the model parameter values with our modelling technique. Of the six fitted model parameters, four are consistent with the truth values within $1\sigma$, and all six within $2\sigma$. We similarly find good agreement between the fitted parameters and the truth values for the static source, with four parameters within $1\sigma$ of the truth value and another within $2\sigma$. The posterior distribution for $\theta$ should be uniform, however, as discussed in Section~\ref{sec:visibility modelling discussion}, this is likely the result of the low signal to noise and poor $uv$-coverage of the synthetic observation. This structure is replicated in the posterior distribution for the model fit to epoch A, allowing us to conclude we can identify the component in epoch A as being stationary. In the example synthetic observations we present here, we note that the full-width half-maximum size of the components is overestimated in our modelling (although they are still within 2$\sigma$). This is likely the result of the component size approaching the resolution limit of these simulated observations at 8.4\,GHz.

\section{\mj\ Model Fit Priors and Posterior Distributions }\label{sec:appendix real}
\begin{figure*}
    \centering
    \includegraphics[width=\linewidth]{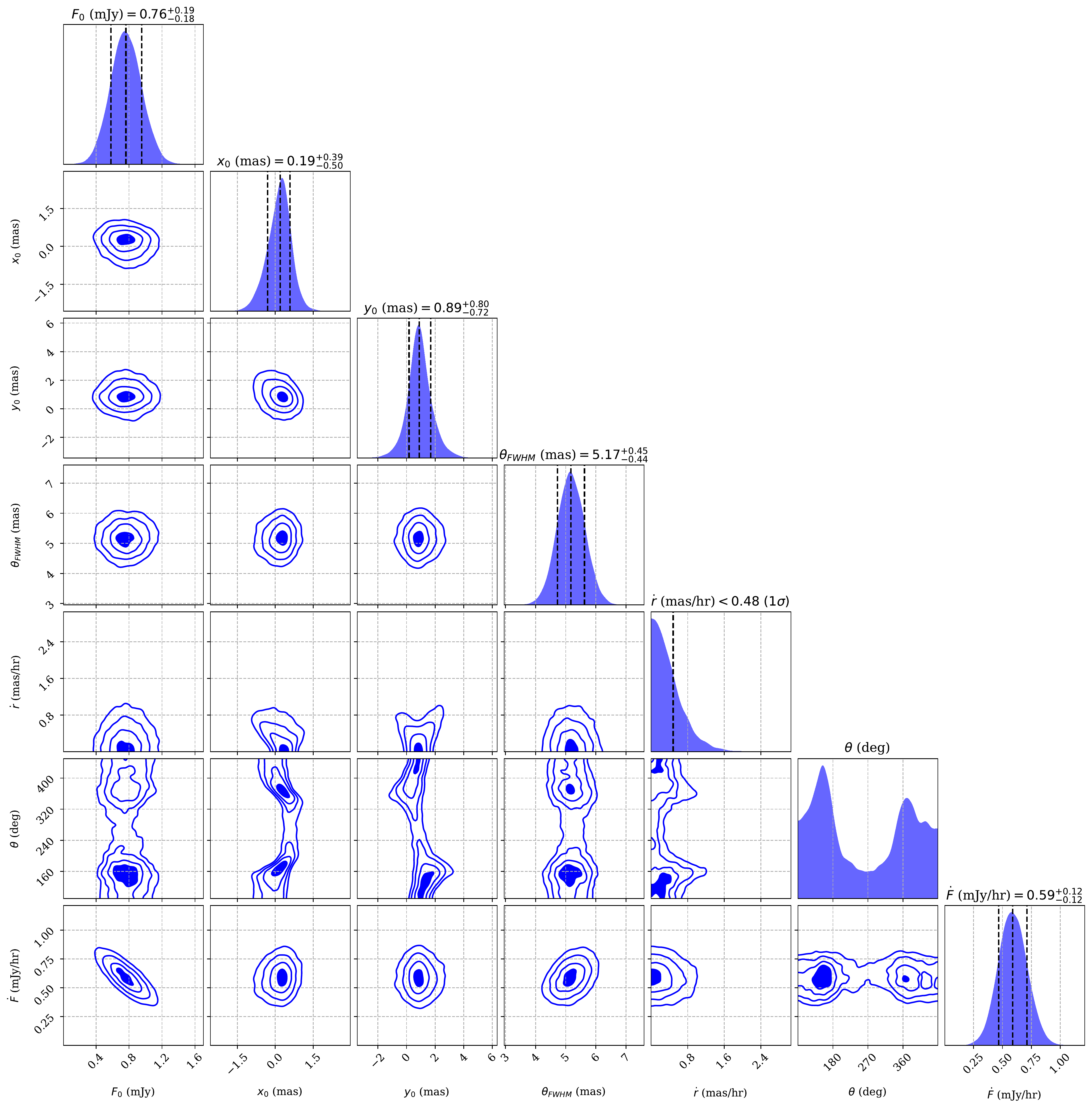}
    \caption{Marginal and joint posterior distributions for the fitted parameters for epoch A. For $\dot{r}$, we mark the 1$\sigma$ upper limit, and for $\theta$ we mark no credible intervals. For all other parameters the black vertical lines mark the 16th, 50th, and 84th quantiles. The blue contours in the joint posterior distributions mark the 0.5, 1, 1.5 and 2$\sigma$ levels.}
    \label{fig:epoch A corner}
\end{figure*}

\begin{figure*}
    \centering
    \includegraphics[width=\linewidth]{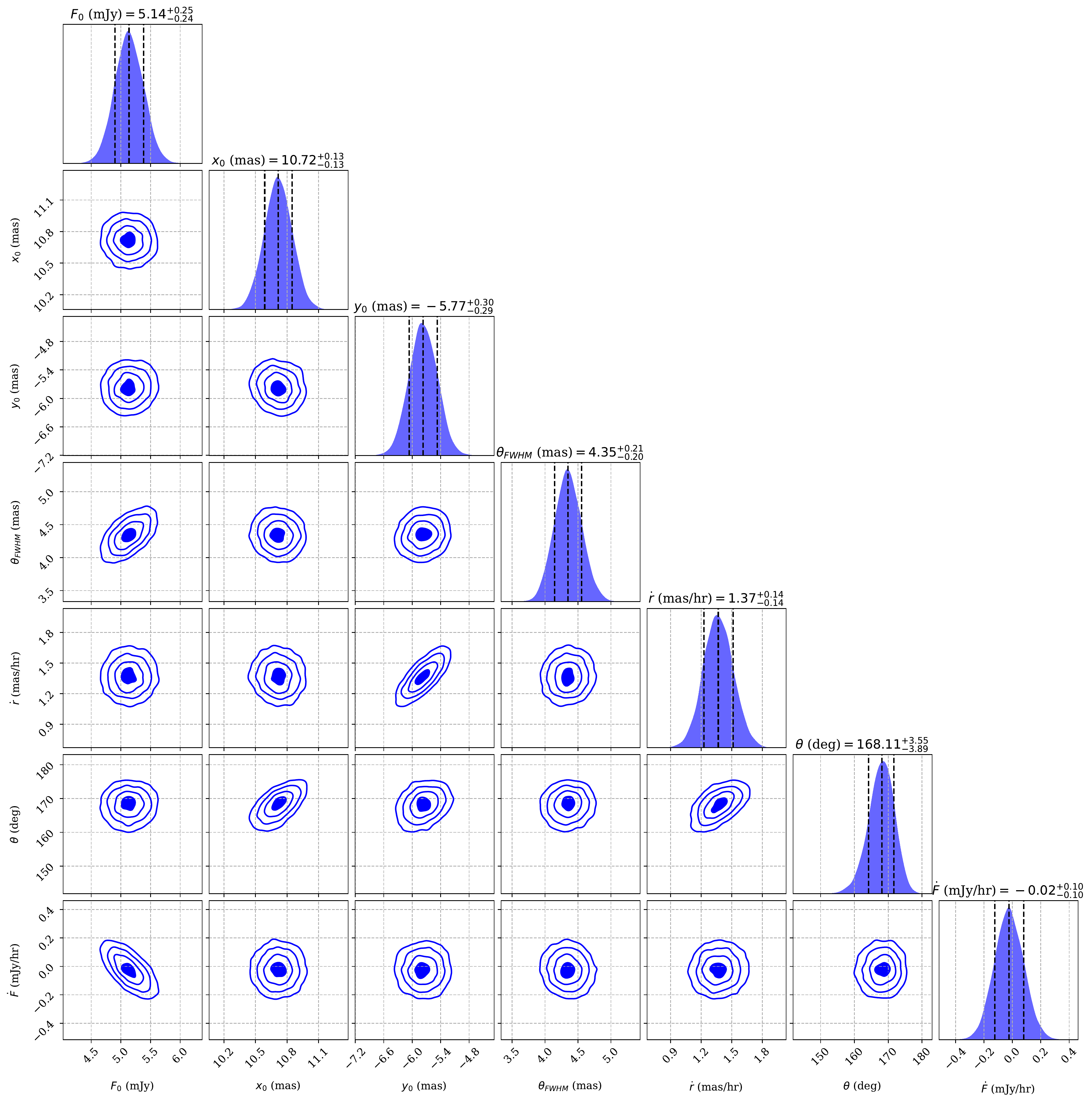}
    \caption{Marginal and joint posterior distributions for the fitted parameters for epoch B. The black vertical lines mark the 16th, 50th, and 84th quantiles. The blue contours in the joint posterior distributions mark the 0.5, 1, 1.5 and 2$\sigma$ levels.}
    \label{fig:epoch B corner}
\end{figure*}

\begin{figure*}
    \centering
    \includegraphics[width=\linewidth]{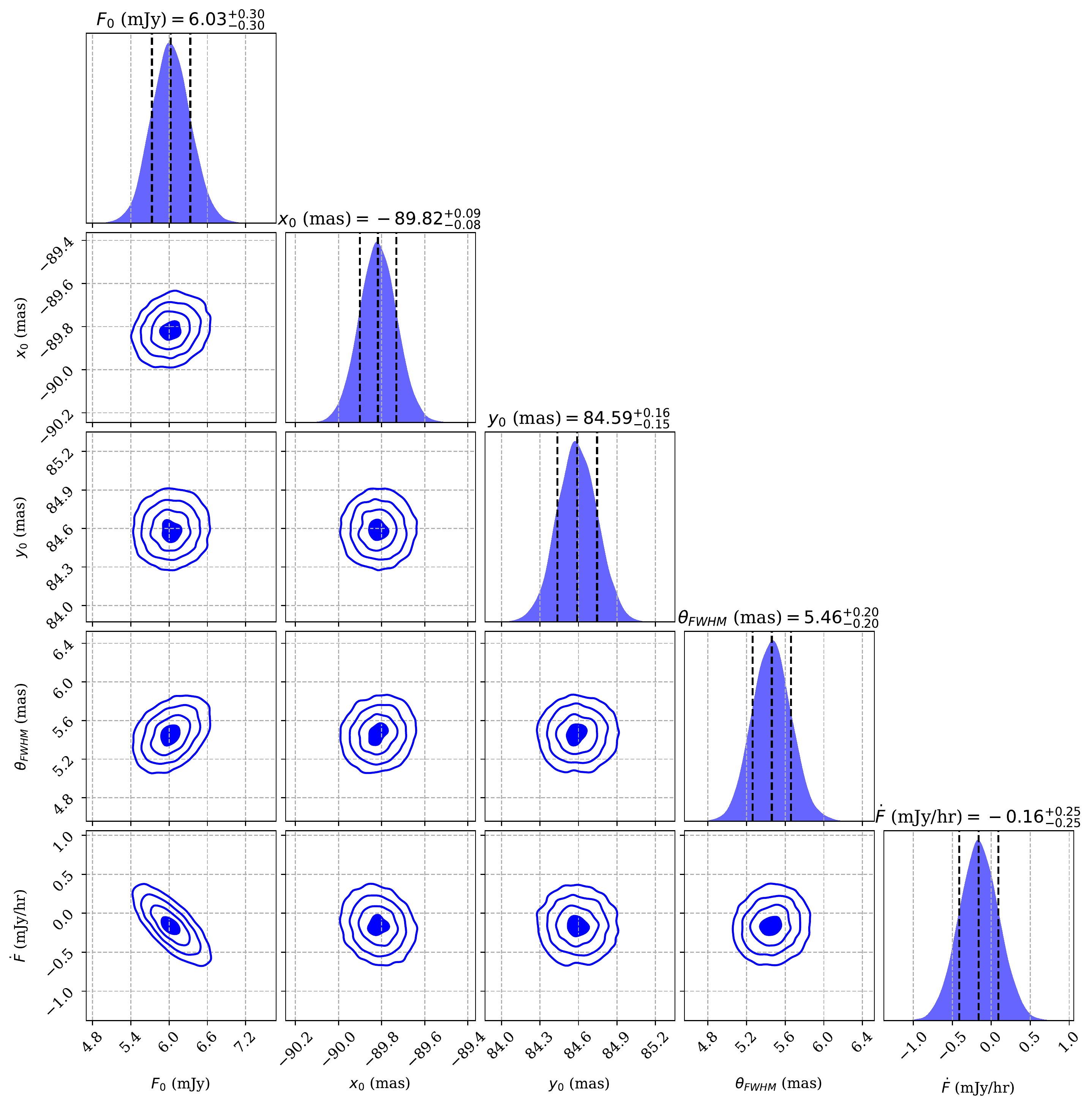}
    \caption{Marginal and joint posterior distributions for the fitted parameters for epoch C. The black vertical lines mark the 16th, 50th, and 84th quantiles. The blue contours in the joint posterior distributions mark the 0.5, 1, 1.5 and 2$\sigma$ levels.}
    \label{fig:epoch C corner}
\end{figure*}

Table~\ref{tab:priors} describes the prior distributions used for our model fits. These priors were based on our initial imaging. For epoch B we placed a relatively tight Gaussian prior on the position angle of the motion of the component, however we were still able to recover the motion of the component when using a uniform prior, confirming that the motion was real and not the result of a poorly chosen prior. In Figs.~\ref{fig:epoch A corner}, \ref{fig:epoch B corner}, and \ref{fig:epoch C corner} we show the joint and marginal probability distributions for our fitted model parameters in epochs A, B, and C, respectively. In epoch A there is clear structure in the posterior distribution for $\dot{r}$ and $\theta$, that closely resembles the structure in those parameters for the synthetic observation with a stationary source. This suggests that the component in epoch A is stationary. 
\begin{table}
    \centering
    \caption{Priors placed on fit parameters for epochs A, B, and C. We used only uniform and normal distributions. For $\dot{F}$, we set the lower boundary for each fit equal to $-F_0/\Delta t$, where $\Delta t$ is the length of the observation, so that the source can never have negative flux density. }
    \begin{tabular}{c|l|l}
        \hline
        Epoch & Parameter           & Prior Distribution                                     \\ \hline \hline
        A     & $F_0$ (mJy)         & $\mathcal{U}(\text{min}=0,\text{max}=10)$              \\
              & $\dot{F}$ (mJy\,\perhour)  & $\mathcal{U}(\text{min}=-F_0/\Delta t,\text{max}=10)$  \\
              & $x_0$ (mas)         & $\mathcal{N}(\mu=0,\sigma=5)$                          \\
              & $y_0$ (mas)         & $\mathcal{N}(\mu=0,\sigma=5)$                          \\
              & $\theta_{\text{FWHM}}$(mas)          & $\mathcal{U}(\text{min}=0,\text{max}=10)$              \\
              & $\dot{r}$ (mas\,\perhour)  & $\mathcal{U}(\text{min}=0,\text{max}=10)$              \\
              & $\theta$ (\degree East of North) & $\mathcal{U}(\text{min}=0,\text{max}=360)$\\ \hline
        B     & $F_0$ (mJy)         & $\mathcal{U}(\text{min}=0,\text{max}=10)$              \\
              & $\dot{F}$ (mJy\,\perhour)  & $\mathcal{U}(\text{min}=-F_0/\Delta t,\text{max}=10)$  \\
              & $x_0$ (mas)         & $\mathcal{N}(\mu=10,\sigma=5)$                         \\
              & $y_0$ (mas)         & $\mathcal{N}(\mu=-5,\sigma=5)$                         \\
              & $\theta_{\text{FWHM}}$(mas)          & $\mathcal{U}(\text{min}=0,\text{max}=10)$              \\
              & $\dot{r}$ (mas\,\perhour)  & $\mathcal{U}(\text{min}=0,\text{max}=10)$              \\
              & $\theta$ (\degree  East of North) & $\mathcal{N}(\mu=135,\sigma=15)$          \\ \hline
        C     & $F_0$ (mJy)         & $\mathcal{U}(\text{min}=0,\text{max}=10)$              \\
              & $\dot{F}$ (mJy\,\perhour)  & $\mathcal{U}(\text{min}=-F_0/\Delta t,\text{max}=10)$  \\
              & $x_0$ (mas)         & $\mathcal{N}(\mu=-85,\sigma=5)$                         \\
              & $y_0$ (mas)         & $\mathcal{N}(\mu=80,\sigma=5)$                         \\
              & $\theta_{\text{FWHM}}$(mas)          & $\mathcal{U}(\text{min}=0,\text{max}=10)$              \\ \hline
    \end{tabular}
    \label{tab:priors}
\end{table}

\section{ATCA Measurements}\label{sec:ATCA Data}
    In Table~\ref{tab:ATCA Data}, we list the subset of ATCA measurements shown in Figs.~\ref{fig:Radio Flare} and \ref{fig:Xray and Radio LC}, which will be presented in full in Espinasse et al. (in prep.).
    
    \begin{table}
        \centering     
        \caption{ATCA measurements of \mj, as presented in Figs.~\ref{fig:Radio Flare} and \ref{fig:Xray and Radio LC}.}
        \begin{tabular}{c|c|c|c}
            \hline
            Date              & Frequency & Flux Density    & $\alpha $       \\ 
            (MJD)             & (GHZ)     & (mJy)           &                 \\
            \hline
            \hline
            $59345.60\pm0.17$ & 5.5       & $1.6\pm0.1$     & $-0.1\pm0.1 $   \\
            $59345.60\pm0.17$ & 9.0       & $1.6\pm0.2$     & $-0.1\pm0.1$    \\
            $59345.60\pm0.17$ & 16.7      & $1.45\pm0.17$   & $-0.1\pm0.1$    \\
            $59345.60\pm0.17$ & 21.2      & $1.3\pm0.3$     & $-0.1\pm0.1$    \\
            $59347.88\pm0.03$ & 5.5       & $9.00\pm0.04$   & $-0.8\pm0.3$    \\
            $59347.88\pm0.03$ & 9.0       & $6.02\pm0.03$   & $-0.8\pm0.3$    \\
            $59351.78\pm0.09$ & 5.5       & $5.30\pm0.02$   & $-0.62\pm0.18$  \\
            $59351.78\pm0.09$ & 9.0       &$ 4.0\pm0.012$   & $-0.62\pm0.18$  \\
            $59353.86\pm0.05$ & 5.5       & $2.03\pm0.05$   & $-0.6\pm0.2$    \\
            $59353.86\pm0.05$ & 9.0       & $1.56\pm0.04$   & $-0.6\pm0.2$    \\
            $59357.88\pm0.03$ & 5.5       & $1.20\pm0.04$   & $-0.40\pm0.15$  \\
            $59357.88\pm0.03$ & 9.0       & $1.00\pm0.03$   & $-0.40\pm0.15$  \\
            $59364.83\pm0.07$ & 5.5       & $0.97\pm0.03$   & $-0.45\pm 0.15$ \\
            $59364.83\pm0.07$ & 9.0       & $0.79\pm0.02$   & $-0.45\pm0.15$  \\
            $59375.85\pm0.02$ & 5.5       & $4.06\pm0.05$   & $-0.47\pm0.14$  \\
            $59375.85\pm0.02$ & 9.0       & $3.27\pm0.03$   & $-0.47\pm 0.14$ \\
            $59382.83\pm0.08$ & 5.5       & $9.31\pm0.03$   & $-0.8\pm0.3$    \\
            $59382.83\pm0.08$ & 9.0       & $6.55\pm0.03$   & $-0.8\pm0.3$    \\
            $59398.59\pm0.03$ & 5.5       & $0.072\pm0.02$  & $-1.0\pm0.9$    \\
            $59398.59\pm0.03$ & 9.0       & $0.045\pm0.015$ & $-1.0\pm0.9$    \\
            \hline 
        \end{tabular}
        \label{tab:ATCA Data}
    \end{table}

%%%%%%%%%%%%%%%%%%%%%%%%%%%%%%%%%%%%%%%%%%%%%%%%%%

% Don't change these lines
\bsp	% typesetting comment
\label{lastpage}
\end{document}